\documentclass[twoside]{mfa}

\usepackage{amssymb,amsfonts}
\renewcommand\volume{{\bf 1} (2012)}        

\usepackage[active]{srcltx}
\usepackage{graphicx}

\newcommand{\pteam}{\pi}

\newcommand{\lteam}{\lambda}

\newcommand{\ppair}{\theta}

\newcommand{\lpair}{\gamma}

\newcommand{\nnum}{n}

\newcommand{\nt}{t}
\newcommand{\ML}[1]{\widehat{#1}}

\newcommand{\pdf}{f}
\newcommand{\pmf}{p}
\newcommand{\prob}{P}
\newcommand{\softmax}{{\boldsymbol{\sigma}}}

\newcommand{\sref}[1]{Section~\ref{#1}}
\newcommand{\aref}[1]{Appendix~\ref{#1}}
\newcommand{\fref}[1]{Figure~\ref{#1}}
\newcommand{\tref}[1]{Table~\ref{#1}}

\DeclareMathOperator{\logistic}{logistic}

\def\commitDATE{ Thu Dec 2 09:10:47 2021 -0500}

 \newcommand{\WLTtieprob}{0.39}
\newcommand{\ECOTprob}{0.38}
 
\numberwithin{equation}{section}

\begin{document}

\title[Bradley-Terry with Multiple Game Outcomes]
{Bradley-Terry Modeling with Multiple Game Outcomes with Applications to
College Hockey}

\author[J.\ T.\ Whelan]{John T.\ Whelan}
\address{School of Mathematical Sciences and Center for Computational Relativity and Gravitation, Rochester Institute of Technology, 85 Lomb Memorial Drive, Rochester, New York 14623, USA \\ and Institute for Theoretical Physics, Goethe University Frankfurt, Max-von-Laue Str.\ 1, D-60438 Frankfurt am Main, Germany}
\email{jtwsma@rit.edu}
\author[J.\ E.\ Klein]{Jacob E.\ Klein}
\address{School of Mathematical Sciences, Rochester Institute of Technology, 85 Lomb Memorial Drive, Rochester, New York 14623, USA}
\email{jek8543@rit.edu}
 \date{\commitDATE}

\keywords{Bradley-Terry, College Hockey, International Hockey}

\subjclass{62F15}{62F07}      

\begin{abstract}
  The Bradley-Terry model has previously been used in both Bayesian
  and frequentist interpretations to evaluate the strengths of sports
  teams based on win-loss game results.  It has also been extended to
  handle additional possible results such as ties.  We implement a
  generalization which includes multiple possible outcomes such as
  wins or losses in regulation, overtime, or shootouts.  A natural
  application is to ice hockey competitions such as international
  matches, European professional leagues, and NCAA hockey, all of
  which use a zero-sum point system which values overtime and shootout
  wins as $2/3$ of a win, and overtime and shootout losses as $1/3$ of
  a win.  We incorporate this into the probability model, and evaluate
  the posterior distributions for the associated strength parameters
  using techniques such as Gaussian expansion about maximum a
  posteriori estimates, and Hamiltonian Monte Carlo.
\end{abstract}

\newtheorem{theorem}{Theorem}[section]
\newtheorem{corollary}[theorem]{Corollary}
\newtheorem{lemma}[theorem]{Lemma}
\newtheorem{proposition}[theorem]{Proposition}

\theoremstyle{definition}
\newtheorem{definition}[theorem]{Definition}
\newtheorem{problem}[theorem]{Problem}
\newtheorem{example}[theorem]{Example}
\newtheorem{remark}[theorem]{Remark}

\numberwithin{equation}{section}

\maketitle

\section{Introduction}

The Bradley-Terry model \cite{BRADLEY01121952,Zermelo1929} has long
been used for evaluating paired comparisons, such as games between
pairs of teams in which one team or the other wins each game.  The
model assigns a strength parameter to each team, and the odds ratio
associated with the probability of a team winning a game is equal to
the ratio of the strengths.  These strength parameters can be
estimated based on the full set of game results and used to rank teams
or make future predictions.  The model has been extended by Davidson
\cite{Davidson1970} to contests in which a tie or drawn contest is a
possible outcome.  For many years, such ties were a common occurrence
in the sport of ice hockey, but recently tie-breaking methods such as
an overtime period played under different rules and/or a shootout in
which the teams alternate penalty shots are used to determine a
winner.  Results in overtime or shootouts can be evaluated differently
from wins in regulation play.  For instance, since 2006\cite{iihf2006}
competitions organized by International Ice Hockey Federation (IIHF) have
awarded three points in the standings to a team winning in regulation,
two points for a win in overtime or a shootout, one point for a loss
in overtime or a shootout, and no points for a loss in regulation, and
many leagues have followed suit.  Compared to the prior system which
awarded two points for a win, one for a tie, and none for a loss,
which effectively treated a tie as half a win and half a loss, the
four-outcome system treats an overtime/shootout win as $2/3$ of a win
and $1/3$ of a loss.\footnote{We do not consider non-zero-sum point
  systems such as that used in association football (soccer) which
  awards three points for a win and one for a draw, so that drawn
  matches are only worth two points total rather than three.
  Likewise, the National Hockey League awards all wins two points and
  overtime/shootout losses one point; this 2-2-1-0 system awards three
  total points for games which go into overtime, but only two for
  games decided in regulation.} One possible approach to either of
these situations (games with three or four outcomes) is to use
standard Bradley-Terry and assign fractional wins as appropriate to the
point system (see, e.g., \cite{Whelan2019}).  However, this is
unsatisfying, as it provides no way to assign a probability for a
future game to end in a tie or overtime or shootout result.  In this paper, we
instead consider a generalization of the tie model of
\cite{Davidson1970} which associates one strength parameter to each
team, along with a single parameter describing the tendency for games
to go into overtime.

The rest of this paper is organized as follows: In \sref{s:models}, we
describe the three models (standard Bradley-Terry,
Bradley-Terry-Davidson including ties, and a new model with four
possible game outcomes including overtime/shootout wins and losses), and
exhibit a generalization of the relevant formulas which describes all
three cases.  In \sref{s:inference} we describe methods for inferring
the relevant parameters of these models given a set of game results:
maximum likelihood estimation, and Bayesian inference using either a
Gaussian approximation or Hamiltonian Monte Carlo.  In \sref{s:demo}
we demonstrate these methods using a recent set
of game results: the
2020-2021 Eastern College Athletic Conference (ECAC) season.
This season
used the standard IIHF system with 3-2-1-0 points assigned for
regulation wins, overtime/shootout wins, overtime/shootout losses, and
regulation losses, respectively.  For the purposes of illustration, we
evaluate the ECAC results with the four-outcome model as well as with
the other two models, treating in one case all wins the same, and in
the other all overtime/shootout results as ties.

\section{Models}
\label{s:models}

In the standard Bradley-Terry model \cite{BRADLEY01121952,Zermelo1929}
each team has a strength $\pteam_i\in(0,\infty)$, and the modelled
probability that team $i$ will win a game with team $j$ is
\begin{equation}
  \ppair^{\text{W}}_{ij} = \frac{\pteam_i}{\pteam_i+\pteam_j}
\end{equation}
so that the probability of a set of game outcomes $D$ in which team
$i$ plays team $j$ $\nnum_{ij}$ times and wins $\nnum^{\text{W}}_{ij}$
of those games is\footnote{The first form explicitly includes each
  pair of teams only once, while the second corrects for the
  double-counting, taking advantage of the fact that
  $n^{\text{W}}_{ii}=0=n^{\text{L}}_{ii}$.
  If the order of the games between pairs of
  teams is ignored, the sampling distribution for the
  $\{\nnum^{\text{W}}_{ij}\}$ is instead
  $\pmf(\{\nnum^{\text{W}}_{ij}\}|\{\pteam_i\}) =
  \left(\prod_{i=1}^{\nt}\prod_{j=1}^{\nt}
  \frac{(\nnum_{ij})!}{(\nnum^{\text{W}}_{ij})!(\nnum^{\text{L}}_{ij})!}
  (\ppair^{\text{W}}_{ij})^{\nnum^{\text{W}}_{ij}}
  (\ppair^{\text{L}}_{ij})^{\nnum^{\text{L}}_{ij}}\right)^{\frac{1}{2}}$.}
\begin{equation}
  \prob(D|\{\pteam_i\})
  = \prod_{i=1}^{\nt}\prod_{j=i+1}^{\nt}
  (\ppair^{\text{W}}_{ij})^{\nnum^{\text{W}}_{ij}}
  (\ppair^{\text{L}}_{ij})^{\nnum^{\text{L}}_{ij}}
  = \left(
    \prod_{i=1}^{\nt}\prod_{j=1}^{\nt}
    (\ppair^{\text{W}}_{ij})^{\nnum^{\text{W}}_{ij}}
    (\ppair^{\text{L}}_{ij})^{\nnum^{\text{L}}_{ij}}
  \right)^{\frac{1}{2}}
  \ ,
\end{equation}
where $\nt$ is the number of teams,
$\nnum^{\text{L}}_{ij}=\nnum^{\text{W}}_{ji}=\nnum_{ij}-\nnum^{\text{W}}_{ij}$
and
$\ppair^{\text{L}}_{ij}=\ppair^{\text{W}}_{ji}=1-\ppair^{\text{W}}_{ij}$.

Davidson \cite{Davidson1970} proposed an extension for competitions
which include the probabilities of ties, in which the probabilities of
the three possible outcomes of a game are
\begin{subequations}
  \begin{align}
    \ppair^{\text{W}}_{ij}
    &= \frac{\pteam_i}{\pteam_i + \nu\sqrt{\pteam_i\pteam_j} + \pteam_j}
    \\
    \ppair^{\text{T}}_{ij}
    &= \frac{\nu\sqrt{\pteam_i\pteam_j}}
      {\pteam_i + \nu\sqrt{\pteam_i\pteam_j} + \pteam_j}
    \\
    \ppair^{\text{L}}_{ij}
    &= \frac{\pteam_j}{\pteam_i + \nu\sqrt{\pteam_i\pteam_j} + \pteam_j}
      \ ,
  \end{align}
\end{subequations}
where $\nu\in[0,\infty)$ is an additional parameter which describes
how likely ties are to occur.  (The probability of a tie in a game
between evenly matched teams is $\frac{\nu}{2+\nu}$.)  Evidently,
$\ppair^{\text{L}}_{ij}=\ppair^{\text{W}}_{ji}$,
$\ppair^{\text{T}}_{ij}=\ppair^{\text{T}}_{ji}$ and
$\ppair^{\text{W}}_{ij}+\ppair^{\text{T}}_{ij}+\ppair^{\text{L}}_{ij}=1$.
The probability of a given set of game outcomes in which the
$\nnum_{ij}=\nnum^{\text{W}}_{ij}+\nnum^{\text{T}}_{ij}+\nnum^{\text{L}}_{ij}$
games between teams $i$ and $j$ result in $\nnum^{\text{W}}_{ij}$
wins, $\nnum^{\text{T}}_{ij}$ ties and $\nnum^{\text{L}}_{ij}$ losses
for team $i$ (where $\nnum^{\text{T}}_{ij}=\nnum^{\text{T}}_{ji}$ and
$\nnum^{\text{L}}_{ij}=\nnum^{\text{W}}_{ji}$) is
\begin{equation}
  \prob(D|\{\pteam_i\},\nu)
  = \left(
    \prod_{i=1}^{\nt}\prod_{j=1}^{\nt}
    (\ppair^{\text{W}}_{ij})^{\nnum^{\text{W}}_{ij}}
    (\ppair^{\text{T}}_{ij})^{\nnum^{\text{T}}_{ij}}
    (\ppair^{\text{L}}_{ij})^{\nnum^{\text{L}}_{ij}}
  \right)^{\frac{1}{2}}
  \ .
\end{equation}

We propose an extension appropriate for a system in which a win in
overtime or a shootout is treated as $2/3$ of a win and $1/3$ of a loss,
Writing the four possible game outcomes as RW for regulation win, OW
for overtime/shootout win, OL for overtime/shootout loss,
and RL for regulation loss, the modelled probability
of each outcome would be\footnote{The exponents are chosen to
  correspond to the share of the points ($2/3$ and $1/3$,
  respectively) awarded for an overtime/shootout win or loss.  This
  has the desirable feature that the maximum likelihood equation
  \eqref{e:pMLE} becomes (after multiplying by $3$)
  $$
  \sum_{j=1}^{\nt} \nnum_{ij}
  \left(
    3\ML{\ppair}^{\text{RW}}_{ij}
    + 2\ML{\ppair}^{\text{OW}}_{ij}
    + \ML{\ppair}^{\text{OL}}_{ij}
  \right)
  =
  \sum_{j=1}^{\nt}
  \left(
    3\nnum^{\text{RW}}_{ij} + 2\nnum^{\text{OW}}_{ij} + \nnum^{\text{OL}}_{ij}
  \right)
  \ ,
  $$
  i.e., that the expected number of points for each team equals the
  actual number.  See also the discussion in \sref{s:conclusions}
  about possible alternative models, including extended models in
  which the exponents are not fixed, but inferred from the data.}
\begin{subequations}
  \begin{align}
    \ppair^{\text{RW}}_{ij}
    &= \frac{\pteam_i}{\pteam_i + \nu\pteam_i^{2/3}\pteam_j^{1/3}
      + \nu\pteam_i^{1/3}\pteam_j^{2/3} + \pteam_j}
    \\
    \ppair^{\text{OW}}_{ij}
    &= \frac{\nu\pteam_i^{2/3}\pteam_j^{1/3}}
      {\pteam_i + \nu\pteam_i^{2/3}\pteam_j^{1/3}
      + \nu\pteam_i^{1/3}\pteam_j^{2/3} + \pteam_j}
    \\
    \ppair^{\text{OL}}_{ij}
    &= \frac{\nu\pteam_i^{1/3}\pteam_j^{2/3}}
      {\pteam_i + \nu\pteam_i^{2/3}\pteam_j^{1/3}
      + \nu\pteam_i^{1/3}\pteam_j^{2/3} + \pteam_j}
    \\
    \ppair^{\text{RL}}_{ij}
    &= \frac{\pteam_j}{\pteam_i + \nu\pteam_i^{2/3}\pteam_j^{1/3}
      + \nu\pteam_i^{1/3}\pteam_j^{2/3} + \pteam_j}
      \ .
  \end{align}
\end{subequations}
The probability for a set of game outcomes will then be
\begin{equation}
  \prob(D|\{\pteam_i\},\nu)
  = \left(
    \prod_{i=1}^{\nt}\prod_{j=1}^{\nt}
    (\ppair^{\text{RW}}_{ij})^{\nnum^{\text{RW}}_{ij}}
    (\ppair^{\text{OW}}_{ij})^{\nnum^{\text{OW}}_{ij}}
    (\ppair^{\text{OL}}_{ij})^{\nnum^{\text{OL}}_{ij}}
    (\ppair^{\text{RL}}_{ij})^{\nnum^{\text{RL}}_{ij}}
  \right)^{\frac{1}{2}}
  \ .
\end{equation}

If we write $\lteam_i=\ln\pteam_i\in(-\infty,\infty)$ and
$\tau=\ln\nu\in(-\infty,\infty)$, we can describe all three models as
special cases of a general model in which the probability of a game
between teams $i$ and $j$ ending in outcome $I$ is
\begin{equation}
  \label{e:thetaImodel}
  \ppair^{I}_{ij}
  = \frac{\pteam_i^{p_I} \pteam_j^{1-p_I} \nu^{o_I}}
  {\sum_J \pteam_i^{p_J} \pteam_j^{1-p_J} \nu^{o_J}}
  = \frac{(\pteam_i/\pteam_j)^{p_I}\nu^{o_I}}
  {\sum_J (\pteam_i/\pteam_j)^{p_J} \nu^{o_J}}
  = \softmax(\{p_J(\lteam_i-\lteam_j)+o_J\tau|J\})_I
  \ ,
\end{equation}
where
\begin{equation}
  \softmax(\mathbf{x})_I = \frac{e^{x_I}}{\sum_J e^{x_J}}
\end{equation}
is a vector equivalent of the logistic function known as the softmax
function.\cite{Bridle1990}  The probability for a set of game outcomes is
\begin{equation}
  \prob(D|\{\pteam_i\},\nu)
  = \left(
    \prod_{i=1}^{\nt}\prod_{j=1}^{\nt}\prod_I
    (\ppair^{I}_{ij})^{\nnum^{I}_{ij}}
  \right)^{\frac{1}{2}}
  \ .
\end{equation}
Specifically,
\begin{itemize}
\item For the standard Bradley-Terry model, $p_{\text{W}}=1$,
  $p_{\text{L}}=0$, and $o_{\text{W}}=o_{\text{L}}=0$.
\item For the Bradley-Terry-Davidson model with ties,
  $p_{\text{W}}=1$, $p_{\text{T}}=\frac{1}{2}$, $p_{\text{L}}=0$,
  $o_{\text{W}}=o_{\text{L}}=0$, and $o_{\text{T}}=1$.
\item For the model introduced in this paper, $p_{\text{RW}}=1$,
  $p_{\text{OW}}=\frac{2}{3}$, $p_{\text{OL}}=\frac{1}{3}$,
  $p_{\text{RL}}=0$, $o_{\text{RW}}=o_{\text{RL}}=0$, and
  $o_{\text{OW}}=o_{\text{OL}}=1$.
\end{itemize}
All of these models satisfy $\sum_I\ppair^{I}_{ij}=1$, and have
``opposite'' outcomes $I$ and $-I$ such that
$\ppair^{-I}_{ij}=\ppair^{I}_{ji}$, $p_{-I}=1-p_{I}$, and
$o_{-I}=o_{I}$.  They also satisfy $0\le p_I \le 1$ and
$o_I\in\{0,1\}$.  We confine ourselves below to cases where these
properties hold.

\section{Inference of Parameters}

\label{s:inference}

\subsection{Maximum Likelihood}

Maximum likelihood estimates (MLEs) of Bradley-Terry strength parameters
\cite{Zermelo1929,Ford:1957,Davidson1970} provide a straightforward
way of associating a ``rating'' to each team based on their game
results, and have been proposed as a replacement for less reliable
ways of evaluating a team's game results in light of the difficulty of
their schedule.\cite{KRACH1993}

We can consider the probability
$\prob(D|\{\pteam_i\},\nu)=\prob(D|\{\lteam_i\},\tau)$ as a likelihood
function of the parameters $\{\lteam_i\}$ and $\tau$, with
log-likelihood
\begin{equation}
  \ln\prob(D|\{\lteam_i\},\tau)
  = \frac{1}{2}
  \sum_{i=1}^{\nt}\sum_{j=1}^{\nt}\sum_I
  \nnum^{I}_{ij}\ln\ppair^{I}_{ij}
  \ .
\end{equation}
We can use the identity
\begin{equation}
  d \ln\softmax(\mathbf{x})_I = dx_I - \sum_{J} dx_J\softmax(\mathbf{x})_J
\end{equation}
to show that
\begin{subequations}
  \begin{equation}
    \frac{\partial\ln\ppair^{I}_{ij}}{\partial\tau}
    = o_I - \sum_J o_J \ppair^{J}_{ij}
  \end{equation}
  and
  \begin{equation}
    \frac{\partial\ln\ppair^{I}_{ij}}{\partial\lteam_k}
= (\delta_{ik}-\delta_{jk}) \left(p_I-\sum_J p_J\ppair^{J}_{ij}\right)
    \ ,
  \end{equation}
\end{subequations}
which means that
\begin{equation}
  \label{e:dlnPdtau}
  \frac{\partial\ln\prob(D|\{\lteam_i\},\tau)}{\partial\tau}
  = \frac{1}{2}\sum_{i=1}^{\nt}\sum_{j=1}^{\nt}\sum_I o_I n^{I}_{ij}
  - \frac{1}{2}\sum_{i=1}^{\nt}\sum_{j=1}^{\nt}n_{ij}\sum_I o_I \ppair^{I}_{ij}
\end{equation}
and
\begin{equation}
  \label{e:dlnPdlambda}
  \frac{\partial\ln\prob(D|\{\lteam_i\},\tau)}{\partial\lteam_k}
  = \sum_{i=1}^{\nt} \sum_I n^{I}_{ki} p_I
  - \sum_{i=1}^{\nt} n_{ki} \sum_I p_I \ppair^{I}_{ki}
  \ .
\end{equation}
Using these, we can write the maximum likelihood equations as
\begin{equation}
  n^o
  = \frac{1}{2}\sum_{i=1}^{\nt}\sum_{j=1}^{\nt}n_{ij}
  \sum_I o_I \ML{\ppair}^{I}_{ij}
  = \frac{1}{2}\sum_{i=1}^{\nt}\sum_{j=1}^{\nt}n_{ij}
  \sum_I o_I \frac{(\ML{\pteam}_i/\ML{\pteam}_j)^{p_I} \ML{\nu}^{o_I}}
  {\sum_J (\ML{\pteam}_i/\ML{\pteam}_j)^{p_J} \ML{\nu}^{o_J}}
\end{equation}
and
\begin{equation}
  \label{e:pMLE}
  p_k
  = \sum_{i=1}^{\nt} n_{ki} \sum_I p_I \ML{\ppair}^{I}_{ki}
  = \sum_{i=1}^{\nt} n_{ki} \sum_I p_I
  \frac{(\ML{\pteam}_k/\ML{\pteam}_i)^{p_I} \ML{\nu}^{o_I}}
  {\sum_J (\ML{\pteam}_k/\ML{\pteam}_i)^{p_J} \ML{\nu}^{o_J}}
\end{equation}
where
\begin{equation}
  n^o = \frac{1}{2}\sum_{i=1}^{\nt}\sum_{j=1}^{\nt}\sum_I o_I n^{I}_{ij}
\end{equation}
can be interpreted in the models considered as the number of games
which are tied or go to overtime, respectively, and
\begin{equation}
  p_k = \sum_{i=1}^{\nt} \sum_I n^{I}_{ki} p_I
\end{equation}
can be seen as the total number of ``points'' for team $i$.  The
maximum likelihood equation set each of these quantities equal to
their expectation values.

We can solve the maximum likelihood equations by a generalization of
the iterative method in \cite{Ford:1957}. writing them
\begin{equation}
  \ML{\nu} = n^o
  \left/
    \left(
      \frac{1}{2}\sum_{i=1}^{\nt}\sum_{j=1}^{\nt}n_{ij}
      \frac{\sum_I o_I (\ML{\pteam}_i/\ML{\pteam}_j)^{p_I}}
      {\sum_J (\ML{\pteam}_i/\ML{\pteam}_j)^{p_J} \ML{\nu}^{o_J}}
    \right)
  \right.
\end{equation}
(where we have used the fact that the only non-zero term in the
numerator has $o_I=1$) and
\begin{equation}
  \ML{\pteam}_k = p_k
  \left/
    \left(
      \sum_{i=1}^{\nt} n_{ki}
      \frac{\sum_I p_I\ML{\pteam}_k^{p_I-1}\ML{\pteam}_i^{-p_I} \ML{\nu}^{o_I}}
      {\sum_J (\ML{\pteam}_k/\ML{\pteam}_i)^{p_J} \ML{\nu}^{o_J}}
    \right)
  \right.
  \ .
\end{equation}
As in the standard Bradley-Terry model, the overall multiplicative
scale of $\ML{\pteam}_k$ is undefined (because $\ppair^{I}_{ij}$ can
be written so that the team strengths appear only in the combination
$\pteam_j/\pteam_i$), so it is necessary to rescale the team strengths
at each iteration to preserve a property such as
$\prod_{i=1}^{\nt} \ML{\pteam}_i=1$.  Beyond that, there are
conditions for the maximum likelihood estimates to be finite and
well-defined, which are explored in e.g.,
\cite{Albert1984,Santner1986,ButlerWhelan}.

\subsection{Bayesian Approach}

It is useful to move beyond maximum likelihood estimates, both to
quantify uncertainty in the model parameters, and to make predictions
about the outcome of future games.  (For instance, \cite{Whelan2019}
proposed simulating future games with probabilites drawn from a
posterior distribution capturing the uncertainty in the strength
parameters, rather than fixed probabilties generated from the MLEs of
those parameters.)

A convenient framework for parameter estimates including uncertainties is
Bayesian inference, which defines the posterior probability density
for the parameters $\{\pteam_i\}$ and $\nu$, or equivalently
$\{\lteam_i\}$ and $\tau$, given a set of game results $D$ and prior
assumptions $I$, as
\begin{equation}
  \pdf(\{\lteam_i\},\tau|D,I)
  = \frac{\prob(D|\{\lteam_i\},\tau)\,\pdf(\{\lteam_i\},\tau|I)}{\prob(D|I)}
  \propto \prob(D|\{\lteam_i\},\tau)\,\pdf(\{\lteam_i\},\tau|I)
  \ .
\end{equation}
A variety of choices can be made for the multivariate prior
distribution on $\{\lteam_i\}$ \cite{Whelan2017} in the Bradley-Terry
model, and likewise for the tie/overtime parameter $\tau$.  For
simplicity, we work in this paper with the improper Haldane
prior\footnote{So named because the marginal prior distribution for
  probabilities such as $\ppair_{ij}$ will follow the Haldane prior
  \cite{Haldane1932,Jeffreys1939}, which is the limit of a
  $\text{Beta}(\alpha,\beta)$ distribution as
  $\alpha,\beta\rightarrow 0$.}
\begin{equation}
  \pdf(\{\lteam_i\},\tau|I_0) = \text{constant}
\end{equation}
which means that the posterior distribution is proportional to the
likelihood:
\begin{equation}
  \pdf(\{\lteam_i\},\tau|D,I_0)
  \propto \prob(D|\{\lteam_i\},\tau)
  \ .
\end{equation}
With this choice of prior, the posterior
probability density will be independent of
the combination $\sum_{i=1}^{\nt}\lteam_i$, but otherwise will be
normalizable under the same circumstances that lead to well-defined
maximum likelihood estimates for the parameters.

\subsubsection{Gaussian Approximation}

One convenient approach is to Taylor expand the log-posterior
$\ln\pdf(\{\lteam_i\},\tau|D,I)$ about the maximum a posteriori
solution (which in this case is the maximum likelihood solution
$\{\ML{\lteam}_i\},\ML{\tau}$).\footnote{Note that this method does
  not assign special significance to the MAP estimates, but uses them
  as the starting point for a convenient approximation to the
  posterior probability distribution.}  Truncating the expansion at
second order gives a Gaussian approximation
\begin{multline}
  \pdf(\{\lteam_i\},\tau|D,I_0) \approx
  \pdf(\{\ML{\lteam}_i\},\tau|D,I_0)
  \exp\left(
    -\frac{1}{2}\sum_{i=1}^{\nt}\sum_{j=1}^{\nt} H_{ij}
    \left(\lteam_i-\ML{\lteam}_i\right)\left(\lteam_j-\ML{\lteam}_j\right)
  \right.
  \\
  \left.
    -\sum_{i=1}^{\nt} H_{i\tau}
    \left(\lteam_i-\ML{\lteam}_i\right)\left(\tau-\ML{\tau}\right)
    -\frac{1}{2} H_{\tau\tau} \left(\tau-\ML{\tau}\right)^2
  \right)
  \ ,
\end{multline}
where $\mathbf{H}$ is the $(\nt+1)\times(\nt+1)$ Hessian matrix
\begin{equation}
  \mathbf{H}
  =
  \begin{pmatrix}
    \{H_{ij}\} & \{H_{i\tau}\} \\
    \{H_{\tau j}\} & H_{\tau\tau} \\
  \end{pmatrix}
\end{equation}
with elements\footnote{Note the
  similarity to the Fisher information matrix
  $I_{ij}(\{\lteam_k\})=\sum_D
  P(D|\{\lteam_k\},I)\frac{\partial^2}{\partial\lteam_i\partial\lteam_j}
  [-\ln P(D|\{\lteam_k\},I)]$, which differs from the Hessian in that
  that $H_{ij}$ depends on the observed data, while $I_{ij}$ is a
  function defined on parameter space.}
\begin{subequations}
  \begin{gather}
    H_{ij} = -\left[
      \frac{\partial^2}{\partial\lteam_i\partial\lteam_j}
      \ln P(D|\{\lteam_k\},\tau)
    \right]_{\{\lteam_k=\ML{\lteam}_k\},\tau=\ML{\tau}}
    \\
    H_{i\tau} = H_{\tau i} = -\left[
      \frac{\partial^2}{\partial\lteam_i\partial\tau}
      \ln P(D|\{\lteam_k\},\tau)
    \right]_{\{\lteam_k=\ML{\lteam}_k\},\tau=\ML{\tau}}
    \\
    H_{\tau\tau} = -\left[
      \frac{\partial^2}{\partial\tau^2}
      \ln P(D|\{\lteam_k\},\tau)
    \right]_{\{\lteam_k=\ML{\lteam}_k\},\tau=\ML{\tau}}
    \ .
  \end{gather}
\end{subequations}
To compute the elements of the Hessian matrix, we return to the first
derivative \eqref{e:dlnPdtau} and differentiate them to get
\begin{equation}
  - \frac{\partial^2\ln\prob(D|\{\lteam_i\},\tau)}{\partial\tau^2}
  = \frac{1}{2}\sum_{i=1}^{\nt}\sum_{j=1}^{\nt}n_{ij}\sum_I o_I
  \frac{\partial \ppair^{I}_{ij}}{\partial\tau}
= \frac{1}{2}\sum_{i=1}^{\nt}\sum_{j=1}^{\nt}n_{ij}
  \ppair^o_{ij}(1-\ppair^o_{ij})
  \ ,
\end{equation}
where
\begin{equation}
  \ppair^o_{ij} = \sum_I o_I \ppair^{I}_{ij}
\end{equation}
is the probability of a tie or overtime game, depending on the model,
and we have used the fact that $o_I^2=o_I$ since $o_I\in\{0,1\}$.
Similarly, using the properties $\sum_I\ppair^{I}_{ij}=1$,
$\ppair^{-I}_{ij}=\ppair^{I}_{ji}$, $p_{-I}=1-p_{I}$, and
$o_{-I}=o_{I}$, we find
\begin{equation}
  - \frac{\partial^2\ln\prob(D|\{\lteam_i\},\tau)}{\partial\tau\partial\lteam_k}
= \sum_{i=1}^{\nt}n_{ki}
  \sum_I o_I \ppair^{I}_{ki}
  \left(
    p_I - \sum_J p_J\ppair^{J}_{ki}
  \right)
\end{equation}
and, finally, differentiating \eqref{e:dlnPdlambda} gives us
\begin{multline}
  - \frac{\partial^2\ln\prob(D|\{\lteam_i\},\tau)}
  {\partial\lteam_k\partial\lteam_\ell}
= \delta_{k\ell} \sum_{i=1}^{\nt} n_{ki} \sum_I p_I\ppair^{I}_{ki}
  \left(p_I-\sum_J p_J\ppair^{J}_{ki}\right)
  \\
  -  n_{k\ell} \sum_I p_I\ppair^{I}_{k\ell}
  \left(p_I-\sum_J p_J\ppair^{J}_{k\ell}\right)
\end{multline}
so that the Hessian matrix has components
\begin{subequations}
  \label{e:Hessian}
  \begin{gather}
    H_{\tau\tau} = \frac{1}{2}\sum_{i=1}^{\nt}\sum_{j=1}^{\nt}n_{ij}
    \sum_I o_I\ML{\ppair}^{I}_{ij}(1-\sum_Jo_J\ML{\ppair}^{J}_{ij})
    \\
    H_{\tau k} = H_{k\tau}
    = \sum_{i=1}^{\nt}n_{ki}
    \sum_I o_I \ML{\ppair}^{I}_{ki}
    \left(
      p_I - \sum_J p_J\ML{\ppair}^{J}_{ki}
    \right)
    \\
    H_{k\ell}
    = \delta_{k\ell} \sum_{i=1}^{\nt} n_{ki} \sum_I p_I\ML{\ppair}^{I}_{ki}
    \left(p_I-\sum_J p_J\ML{\ppair}^{J}_{ki}\right)
    -  n_{k\ell} \sum_I p_I\ML{\ppair}^{I}_{k\ell}
    \left(p_I-\sum_J p_J\ML{\ppair}^{J}_{k\ell}\right)
    \ .
  \end{gather}
\end{subequations}

Note that in the case of the Bradley-Terry model, where the only
outcomes are win and loss, the condition $o^I=0$ simplifies the
Hessian to $H_{\tau\tau}=H_{\tau k}=0$ (since the $\tau$ parameter is
not actually part of the likelihood), and
\begin{equation}
  H_{k\ell} = \delta_{k\ell} \sum_{i=1}^{\nt} n_{ki}\ML{\ppair}^{\text{W}}_{ki}
  \left(1-\ML{\ppair}^{\text{W}}_{ki}\right)
  -  n_{k\ell}\ML{\ppair}^{\text{W}}_{k\ell}
  \left(1-\ML{\ppair}^{\text{W}}_{k\ell}\right)
\end{equation}
which is the form seen in, e.g., \cite{Whelan2019}.

The Hessian matrix in \eqref{e:Hessian} is singular, since
$\sum_{\ell=1}^{\nt} H_{\tau\ell}=0$ and
$\sum_{\ell=1}^{\nt} H_{k\ell}=0$, which ultimately arise from the
fact that the probabilities $\{\ppair^{I}_{ij}\}$, and thus the
likelihood, are unchanged by adding the same constant to all the
$\{\lteam_i\}$.  This can be handled computationally by constructing a
variance-covariance matrix $\boldsymbol{\Sigma}=\mathbf{H}^+$ which is
the Moore-Penrose pseudo-inverse\cite{penrose_1955}\footnote{For a
  real symmetric matrix with a complete eigenvalue decomposition, this
  operation replaces each non-zero eigenvalue with its reciprocal
  while leaving zero eigenvalues unchanged.} of the Hessian matrix,
and approximating the posterior as a multivariate Gaussian with a mean
of $\{\ML{\lteam}_i\},\ML{\tau}$ and a variance-covariance matrix
$\boldsymbol{\Sigma}$.  This has the effect of enforcing the
constraint
\begin{gather}
  \sum_{i=1}^{\nt} \lteam_i = \sum_{i=1}^{\nt} \ML{\lteam}_i
  = 0
\end{gather}
on the combination of the parameters which has no influence on the
model.

This Gaussian approximation can be used to produce analytic estimates
of quantities of interest, or used for Monte Carlo sampling, as
illustrated in \sref{s:demo}.  It can also be used as a starting point
for importance sampling of the sort discussed in \cite{Whelan2019}.
For the present work, we consider a different Monte Carlo method for
sampling from the exact posterior.

\subsubsection{Hamiltonian Monte Carlo}

Markov-chain Monte Carlo methods provide a convenient way to draw
samples from a posterior distribution.  We demonstrate in this paper
how to draw posterior samples for the Bradley-Terry extensions
considered, using Hamiltonian Monte Carlo as implemented in the Stan
library.\cite{Stan} There are a few technical considerations.  Because
the posterior on $\{\lteam_i\}$ and $\tau$ is improper, trying to draw
from it directly will lead to chains which never converge.  Any
probabilities constructed from the samples will be well-behaved, since
only the meaningless degree of freedom $\sum_{i=1}^{\nt} \lteam_i$ is
unconstrained, but these apparent errors make it more difficult to
detect other potential problems.  It is thus useful instead to
consider only variables $\lpair_{ij}=\lteam_i-\lteam_j$ (and $\tau$)
which contribute to the probability model via (see \eqref{e:thetaImodel})
\begin{equation}
  \ppair^{I}_{ij} = \softmax(\{p_J\lpair_{ij}+o_J\tau|J\})_I
  \ .
\end{equation}
Of course, the full set of $\frac{\nt(\nt-1)}{2}$ values $\lpair_{ij}$
are not independent.  Instead, they are determined by the $\nt-1$
parameters $\omega_i=\lteam_i-\lteam_{i+1}$ for $i=1,\ldots,\nt-1$.
Given the $\{\omega_i\}$ we can construct
$\lpair_{ij} = \sum_{k=i}^{j-1} \omega_k$.

In \aref{s:stanmodel} we show the code of the Stan model used to
perform Hamiltonian Monte Carlo simulations of all three models.

\begin{table}[t!]
  \centering
  \caption{Results of the 2020-2021 Eastern College Athletic
    Conference (ECAC) season, showing the number of regulation wins
    $n^{\text{RW}}_{ij}$ and overtime/shootout wins
    $n^{\text{OW}}_{ij}$ for each team against each opponent.  From
    these we can derive the total number of results of each type (RW,
    OW, OL and RL) for each team, which are used, for example, to
    generate the standings in the 3-2-1-0 point system}
  \begin{tabular}{l| cccc |cccc}
& \multicolumn{4}{|c|}{$\nnum^{\text{RW}}_{ij}$ ($\nnum^{\text{OW}}_{ij}$)}& \multicolumn{4}{|c}{$\nnum^{I}_{i}=\sum_{j=1}^{\nt}n^{I}_{ij}$} \\
Team $i$ & Cg & Ck & Qn & SL& RW & OW & OL & RL
 \\ 
 \hline 
Colgate (Cg)
 & --- & 1(1) & 1(0) & 2(1) & 4 & 2 & 3 & 9 \\ 
Clarkson (Ck)
 & 3(1) & --- & 1(2) & 1(0) & 5 & 3 & 4 & 2 \\ 
Quinnipiac (Qn)
 & 4(1) & 1(2) & --- & 4(1) & 9 & 4 & 2 & 3 \\ 
St.~Lawrence (SL)
 & 2(1) & 0(1) & 1(0) & --- & 3 & 2 & 2 & 7 \\ 
\end{tabular}
   \label{t:ECdata}
\end{table}

\section{Demonstration Using Game Results}

\label{s:demo}

We now illustrate the application of the models described in this
paper using game results from
a competition
which used the 3-2-1-0
point system: the 2020-2021 Eastern College Athletic Conference (ECAC)
season.  While
the league ordinarily plays a balanced round-robin schedule in which
each team plays each other team the same number of times, the season in question ended up being unbalanced due to cancellations of games
arising from the COVID-19 pandemic.  In \tref{t:ECdata} we show the
results for the ECAC season, in the form of $n^{\text{RW}}_{ij}$ and
$n^{\text{OW}}_{ij}$ for each team against each opponent, along with
the total number of results of each type for each team,
$n^I_{i}=\sum_{j=1}^{\nt}n^I_{ij}$.

\subsection{ECAC: Standard Bradley-Terry Model}

\label{s:ECACBT}

As a first demonstration, we consider the standard Bradley-Terry model
applied to the ECAC results with regulation and overtime/shootout wins
being counted as simply ``wins'' and regulation and overtime/shootout
losses being counted as ``losses''.  I.e., we define
$n^{\text{W}}_{ij}=n^{\text{RW}}_{ij}+n^{\text{OW}}_{ij}$ and
$n^{\text{L}}_{ij}=n^{\text{RL}}_{ij}+n^{\text{OL}}_{ij}$.  The
resulting maximum-likelihood solutions $\{\ML{\lteam}_i\}$ and
associated probabilities $\{\ML{\ppair}^{\text{W}}_{ij}\}$ are shown
in \tref{t:ECBTGauss}, along with the uncertainties and correlations
encoded in the variance-covariance matrix $\{\Sigma_{ij}\}$ of the
Gaussian approximation to the posterior distribution.
\begin{table}[t!]
  \centering
  \caption{The maximum likelihood estimates and parameters of the the
    Gaussian approximation to the posterior distribution for the
    Bradley-Terry model applied to the 2020-2021 ECAC results, with
    regulation and overtime/shootout results counted the same.  The
    maximum likelihood estimate $\ML{\lteam}_i$ for each team's
    log-strength has an associated one-sigma uncertainty
    $\sqrt{\Sigma_{ii}}$.  The variance-covariance matrix
    $\{\Sigma_{ij}\}$ can be converted to a correlation matrix
    $\rho_{ij}=\Sigma_{ij}/\sqrt{\Sigma_{ii}\Sigma_{jj}}$.  Note that
    the information included in $\Sigma_{ij}$ is also influenced by
    the constraint $\sum_{i=1}^{\nt}\lteam_i=0$, so for example the
    anti-correlation of the different log-strengths is somewhat
    artificial.  We also show the maximum-likelihood estimates
    $\{\ML{\ppair}^{\text{W}}_{ij}\}$ for the head-to-head win
    probabilities between pairs of teams}
\begin{tabular}{l|cc|cccc|cccc}
& & & \multicolumn{4}{|c}{$\ML{\ppair}^{\text{W}}_{ij}$}& \multicolumn{4}{|c}{$\rho_{ij}=\Sigma_{ij}/\sqrt{\Sigma_{ii}\Sigma_{jj}}$} \\ 
Team & $\ML{\lteam_i}$ & $\sqrt{\Sigma_{ii}}$ & Cg & Ck & Qn & SL & Cg & Ck & Qn & SL \\ 
 \hline 
Cg & $-0.55$ & $0.39$
 & ---  & $0.29$ & $0.22$ & $0.49$ & $1.00$ & $-0.31$ & $-0.39$ & $-0.21$ \\ 
Ck & $0.32$ & $0.43$
 & $0.71$ & ---  & $0.40$ & $0.70$ & $-0.31$ & $1.00$ & $-0.22$ & $-0.50$ \\ 
Qn & $0.74$ & $0.40$
 & $0.78$ & $0.60$ & ---  & $0.78$ & $-0.39$ & $-0.22$ & $1.00$ & $-0.35$ \\ 
SL & $-0.51$ & $0.45$
 & $0.51$ & $0.30$ & $0.22$ & ---  & $-0.21$ & $-0.50$ & $-0.35$ & $1.00$ \\ 
\end{tabular}
   \label{t:ECBTGauss}
\end{table}

\begin{figure}[t!]
  \centering
  \includegraphics[width=0.45\textwidth]{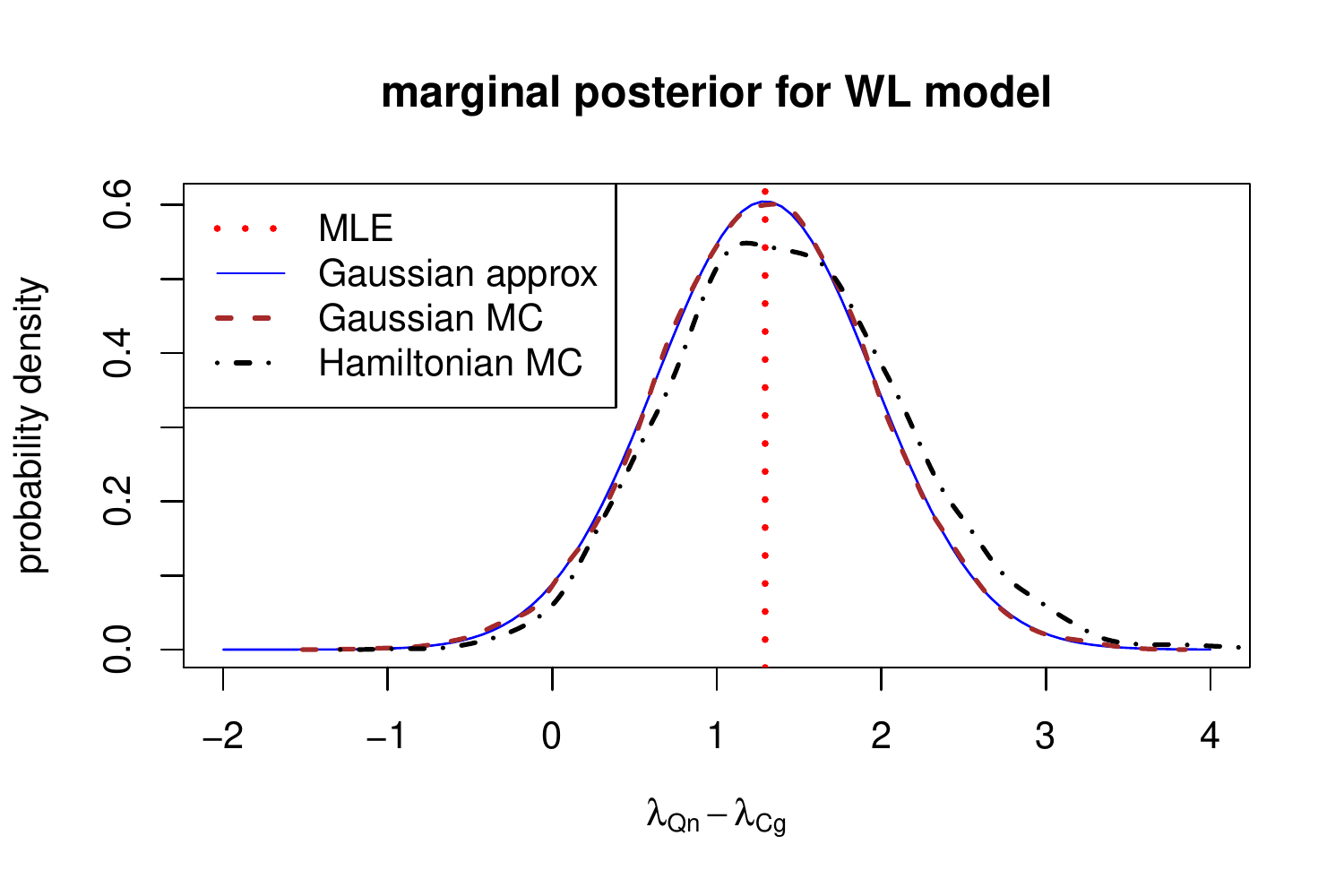}
  \includegraphics[width=0.45\textwidth]{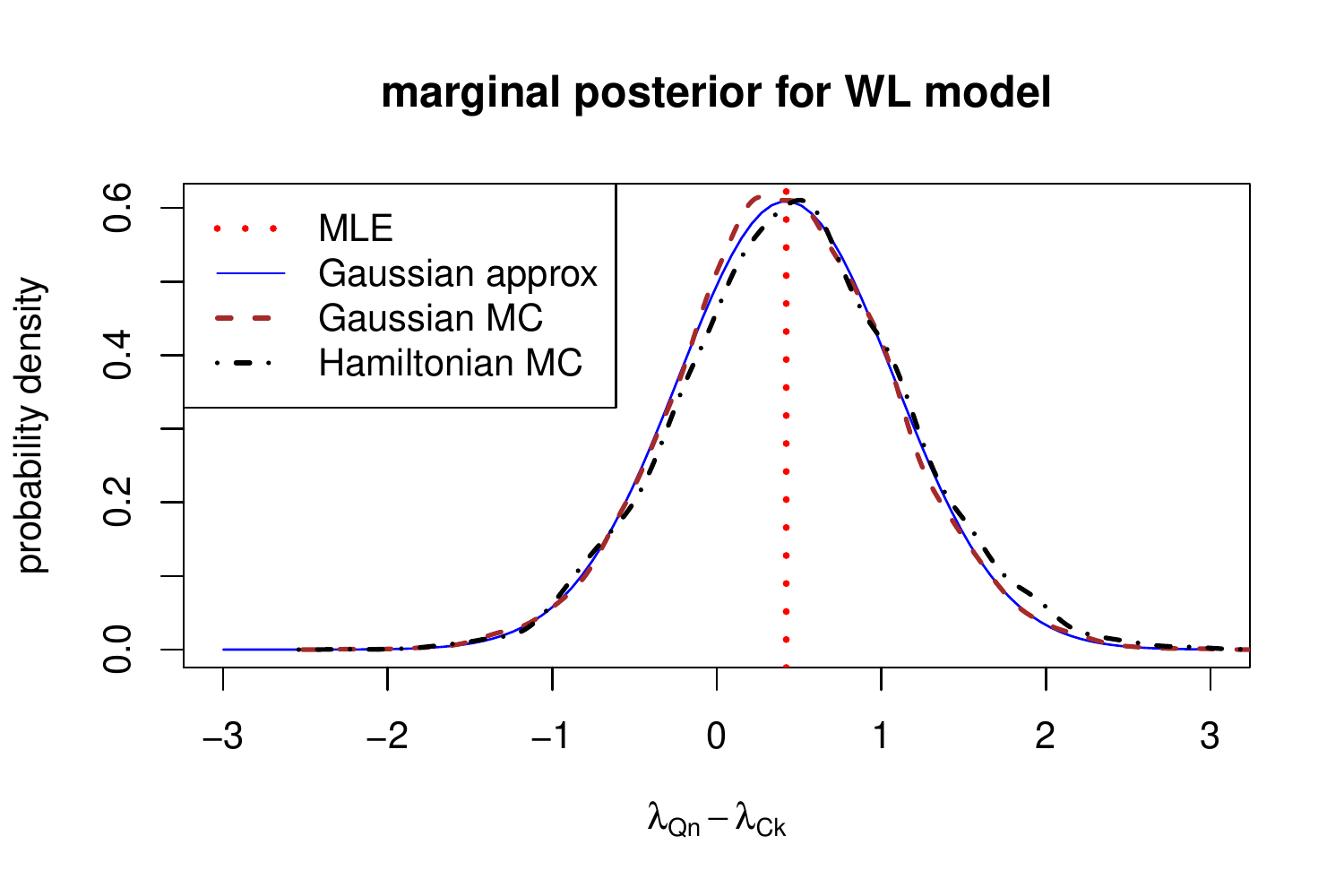}
  \caption{Posterior probability density for difference in
    log-strengths $\lpair_{ij}=\lteam_i-\lteam_j$ between selected
    pairs of teams (left: Quinnipiac and Colgate; right: Quinnipiac
    and Clarkson), based on 2020-2021 ECAC game results in the
    standard Bradley-Terry model with regulation and overtime/shootout
    wins treated the same.  The dotted red vertical line shows the
    maximum likelihod estimate $\ML{\lpair}_{ij}$.  Since the Haldane
    prior used is uniform in the $\{\lteam_i\}$, this is also the
    maximum a posteriori (MAP) value.  The curves show the approximate
    Gaussian posterior from expanding about the MAP value (solid blue
    line), along with density estimates from a set of Monte Carlo
    samples drawn from that distribution (dashed brown line), and a
    set of samples drawn from the exact distribution using Hamiltonian
    Monte Carlo (dot-dash black line).  Differences between the
    Gaussian approximation and the samples from the exact posterior
    are small, but can be noticeable, especially if the maximum
    likelihood estimate $\ML{\lpair}_{ij}$ is far from zero.  For
    reference, note that the ``Gaussian approx'' and ``Gaussian MC''
    curves should only differ due to Monte Carlo errors in the
    construction of the latter}
  \label{f:ECBTQnCg}
\end{figure}
\begin{figure}[t!]
  \centering
  \includegraphics[width=0.45\textwidth]{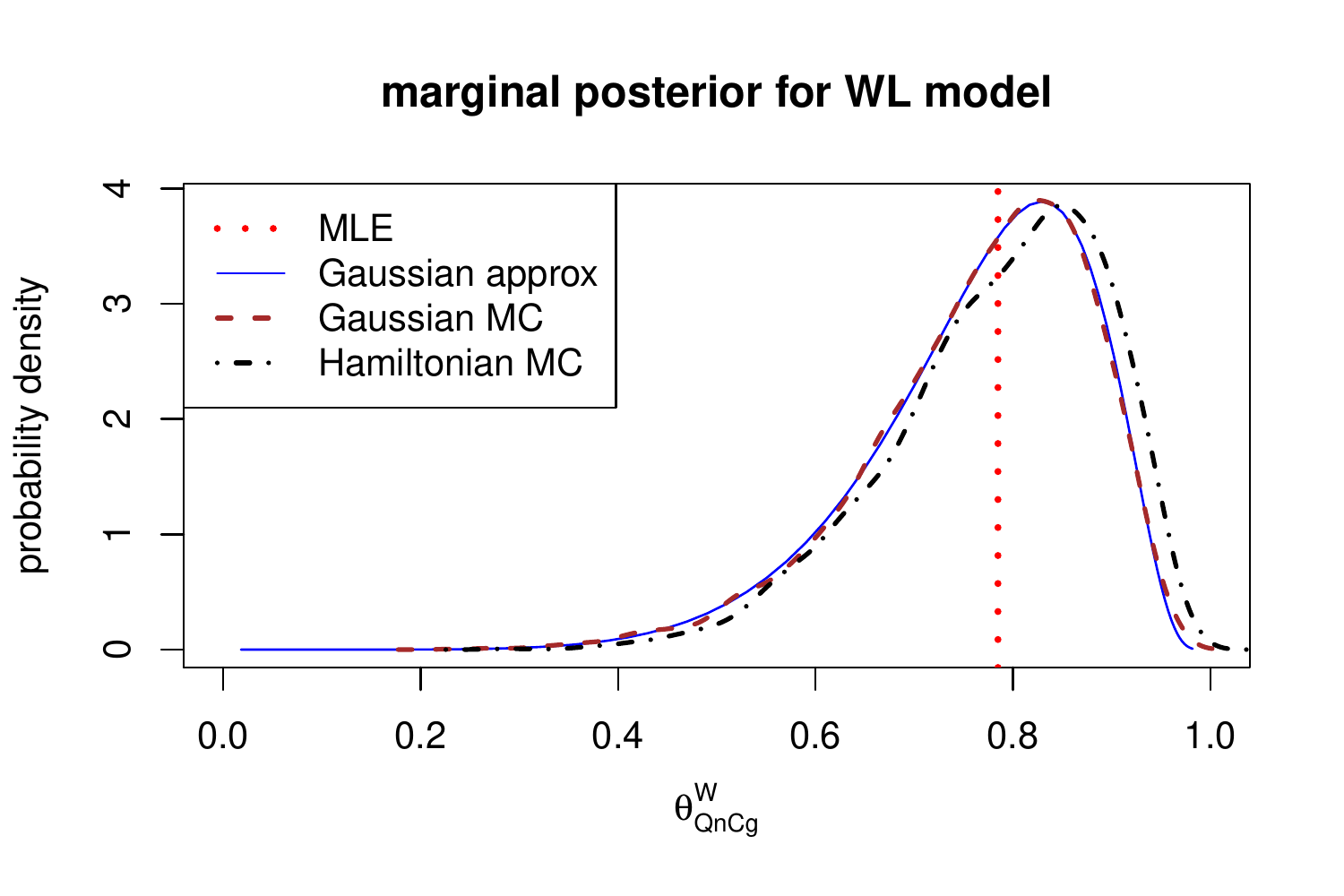}
  \includegraphics[width=0.45\textwidth]{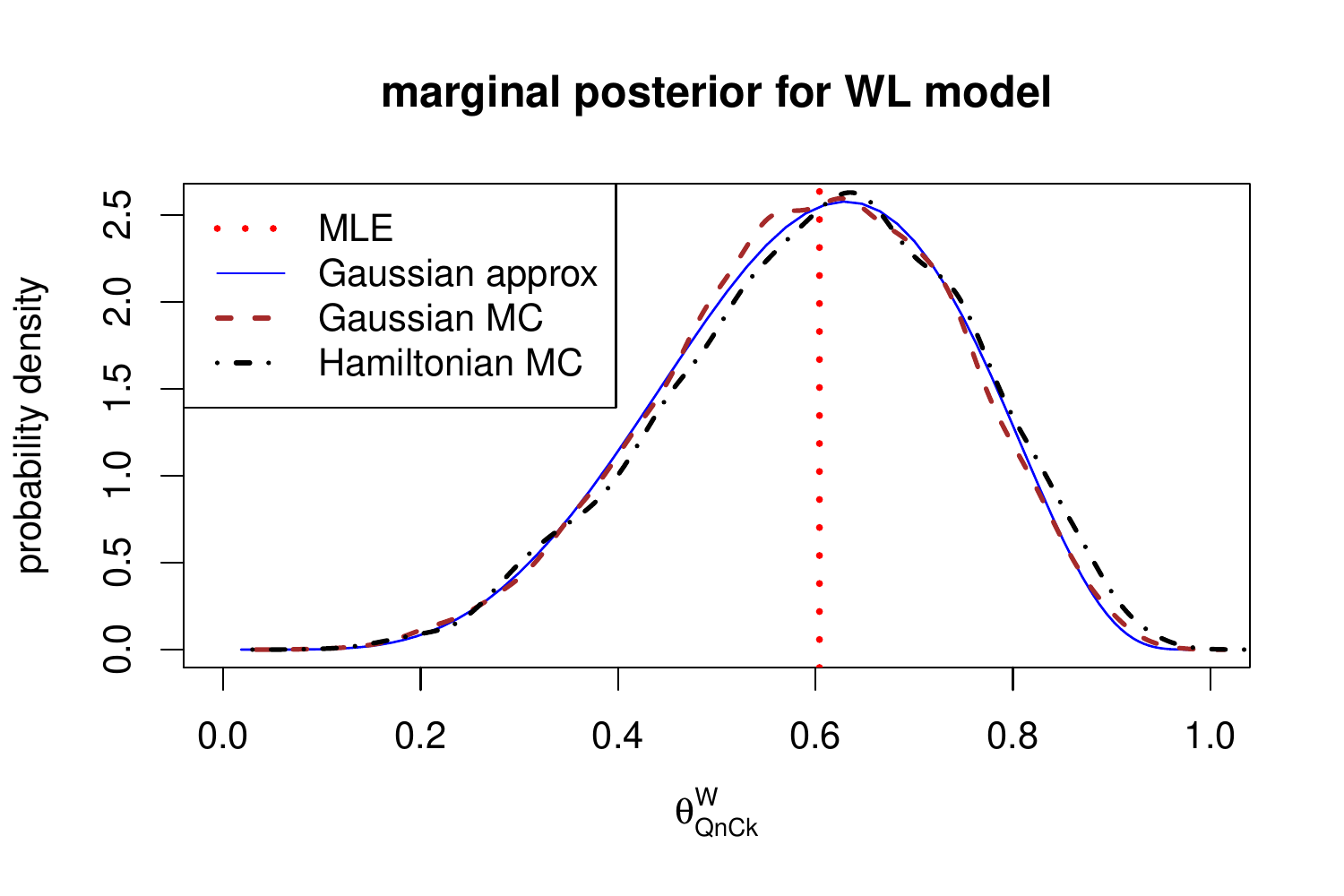}
  \caption{Posterior probability density for the win probability
    $\ppair^{\text{W}}_{ij}=\logistic(\lpair_{ij})$ predicted by the
    Bradley-Terry model for selected pairs of teams, as in
    \fref{f:ECBTQnCg}.  Note that, due to the transformation of the
    probability density, the maximum of the probability density in
    this parameter is not the maximum likelihood value as it was in
    \fref{f:ECBTQnCg}}
  \label{f:ECBTH2HQnCg}
\end{figure}
Since the log-strengths $\{\lteam_i\}$ have an arbitrary additive
scale, a more meaningful understanding of the posterior distributions
is obtained by considering the marginal distribution of the difference
of a pair of team strengths $\lpair_{ij}=\lteam_i-\lteam_j$.  In
\fref{f:ECBTQnCg}, we illustrate the maximum likelihood estimate and
posterior distribution of this
quantity for two of the six pairs of teams: Quinnipiac-Colgate and
Quinnipiac-Clarkson.  We show the posterior in Gaussian approximation (for
which the marginal posterior on $\lpair_{ij}$ is also a Gaussian), in a Monte
Carlo drawn from the approximate multivariate Gaussian distribution,
and in posterior samples from the exact posterior generated using
Hamiltonian Monte Carlo with the Stan library.\cite{Stan}
We can transform the posterior on a difference $\lpair_{ij}$ in
log-strength into a posterior on the corresponding probability
$\ppair^{\text{W}}_{ij}=\logistic(\lpair_{ij})$; this is shown in
\fref{f:ECBTH2HQnCg} for the two sets of posterior samples.
In all cases, the
exact marginal posterior, as estimated by the Hamiltonian Monte Carlo
is only slightly different from the Gaussian approximation.  This is
similar to results found using importance sampling in
\cite{Whelan2019}.

\begin{table}[t!]
  \centering
  \caption{The maximum likelihood estimates for the
    Bradley-Terry-Davidson model applied to the 2020-2021 ECAC
    results, with all overtime games counted as ties.  The maximum
    likelihood estimates $\{\ML{\lteam_i}\}$ and $\ML{\tau}$ of the
    log-strengths and log tie parameter are used to compute the
    estimated probability $\ML{\ppair}^{\text{W}}_{ij}$ for a win and
    $\ML{\ppair}^{\text{T}}_{ij}$ for a tie between each pair of
    teams.  Note that the estimated probability of a game between
    evenly-matched teams to end in a tie is
    $\frac{e^{\ML{\tau}}}{2+e^{\ML{\tau}}}={\WLTtieprob}$, and it is lower the
    more different the two teams' strengths are}
  \begin{tabular}{l|c|cccc}
& & \multicolumn{4}{|c}{$\ML{\ppair}^{\text{W}}_{ij}$ ($\ML{\ppair}^{\text{T}}_{ij}$)} \\ 
Team $i$ & $\ML{\lteam_i}$ & Cg & Ck & Qn & SL \\ 
 \hline 
Cg & $-0.73$
 & ---  & $0.13$ ($0.33$) & $0.11$ ($0.32$) & $0.33$ ($0.38$) \\ 
Ck & $0.70$
 & $0.54$ ($0.33$) & ---  & $0.28$ ($0.38$) & $0.56$ ($0.32$) \\ 
Qn & $0.89$
 & $0.57$ ($0.32$) & $0.34$ ($0.38$) & ---  & $0.59$ ($0.31$) \\ 
SL & $-0.85$
 & $0.29$ ($0.38$) & $0.12$ ($0.32$) & $0.10$ ($0.31$) & ---  \\ 
 \hline 
\multicolumn{6}{c}{$\ML{\tau}=0.23$}\end{tabular}
   \label{t:WLTMLE}
\end{table}
\begin{table}[t!]
  \centering
  \caption{The parameters of the the Gaussian approximation to the
    posterior distribution for the Bradley-Terry-Davidson model
    applied to the 2020-2021 ECAC results, with all overtime games
    counted as ties.  In addition to the log-strength parameters
    considered for the Bradley-Terry model in \tref{t:ECBTGauss}, there
    are uncertainties and correlations associated with the log-tie
    parameter $\tau$}
  \begin{tabular}{l|cc|ccccc}
& & & \multicolumn{4}{|c}{$\rho_{ij}=\Sigma_{ij}/\sqrt{\Sigma_{ii}\Sigma_{jj}}$} \\ 
Team $i$ & $\ML{\lteam}_i$ & $\sqrt{\Sigma_{ii}}$ & Cg & Ck & Qn & SL & $\tau$ \\ 
 \hline 
Cg & $-0.73$ & $0.50$
 & $1.00$ & $-0.35$ & $-0.40$ & $-0.16$ & $-0.22$ \\ 
Ck & $0.70$ & $0.57$
 & $-0.35$ & $1.00$ & $-0.16$ & $-0.53$ & $0.19$ \\ 
Qn & $0.89$ & $0.51$
 & $-0.40$ & $-0.16$ & $1.00$ & $-0.38$ & $0.26$ \\ 
SL & $-0.85$ & $0.58$
 & $-0.16$ & $-0.53$ & $-0.38$ & $1.00$ & $-0.22$ \\ 
$\tau$ & $0.23$ & $0.40$
 & $-0.22$ & $0.19$ & $0.26$ & $-0.22$ & $1.00$ \\ 
\end{tabular}
   \label{t:WLTGauss}
\end{table}

\subsection{ECAC: Bradley-Terry-Davidson Model with Ties}

\label{s:ECACWLT}

Moving on to the Bradley-Terry-Davidson model with ties, we now
consider inference of the log-strength parameters $\{\lteam_i\}$ along
with the log-tie parameter $\tau$.  We illustrate the methods by
reanalyzing the 2020-2021 ECAC results, with all overtime games
treated as ties, so that now $n^{\text{W}}_{ij}=n^{\text{RW}}_{ij}$,
$n^{\text{T}}_{ij}=n^{\text{OW}}_{ij}+n^{\text{OL}}_{ij}$, and
$n^{\text{L}}_{ij}=n^{\text{RL}}_{ij}$.  The maximum likelihood
solutions $\{\ML{\lteam}_i\}$ and $\ML{\tau}$ are shown in
\tref{t:WLTMLE}, along with the associated probabilities
$\{\ML{\ppair}^{\text{W}}_{ij}\}$ for a win and
$\{\ML{\ppair}^{\text{T}}_{ij}\}$ for a tie in contests between pairs
of teams.  In \tref{t:WLTGauss}, we show the maxumum-likelihood
estimates along with the uncertainties in and correlations among the
log-strengths $\{\lteam_i\}$ and the log-tie parameter $\tau$, which
are encoded in the variance-covariance matrix
$\{\Sigma_{ij},\Sigma_{i\tau},\Sigma_{\tau\tau}\}$ of the Gaussian
approximation to the posterior distribution.

\begin{figure}[t!]
  \centering
  \includegraphics[width=0.45\textwidth]{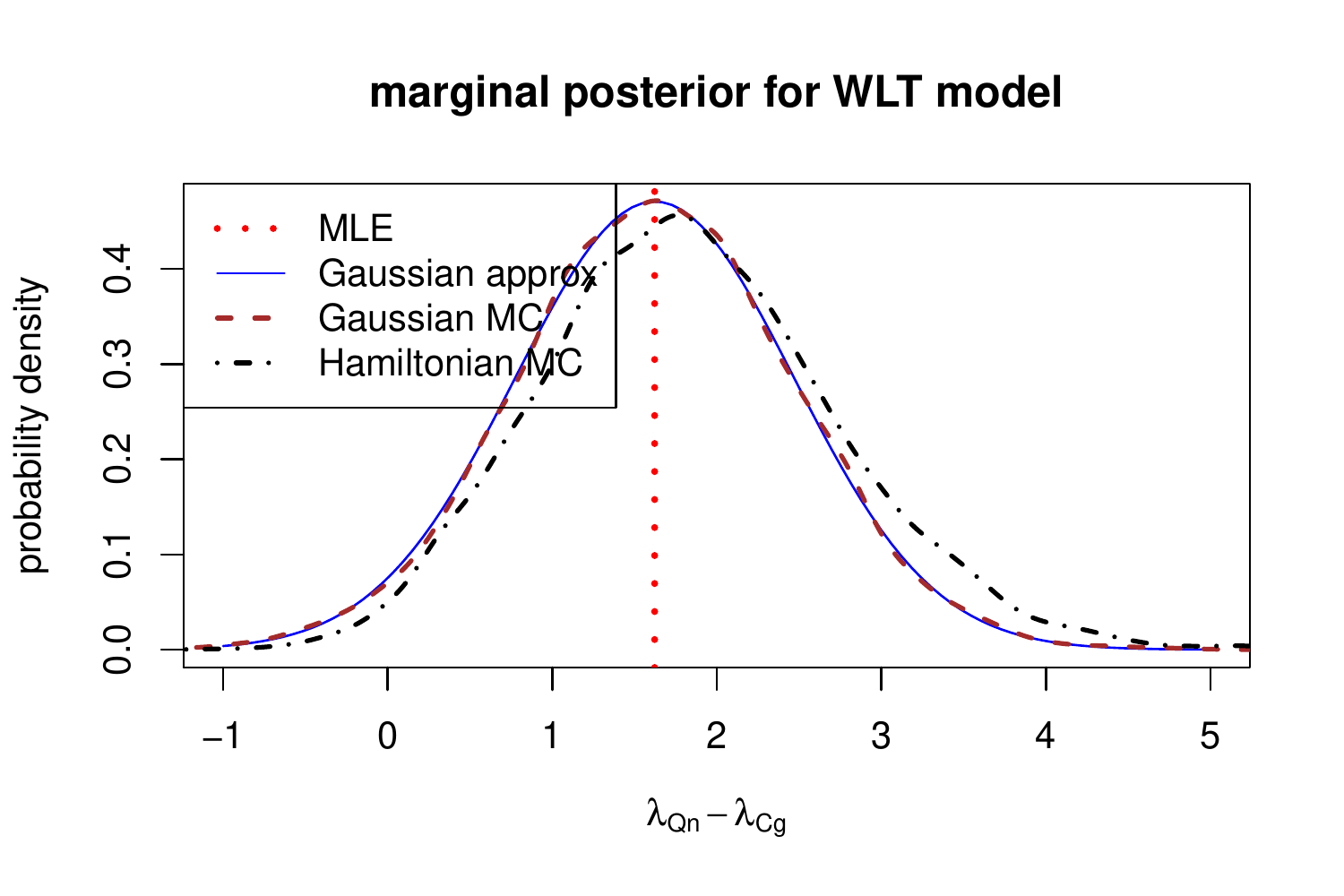}
  \includegraphics[width=0.45\textwidth]{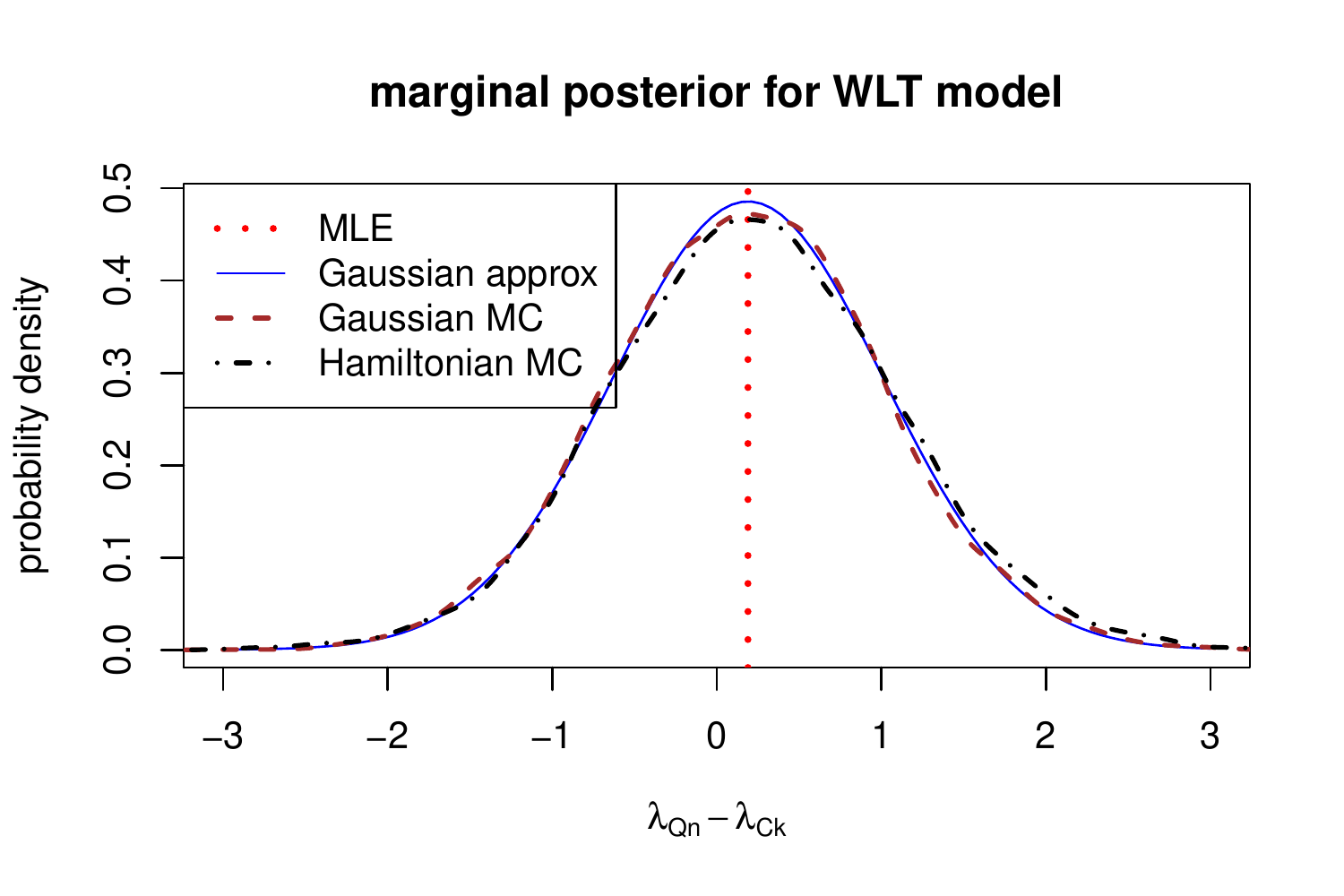}
  \caption{Posterior probability density for difference in
    log-strengths $\lteam_i-\lteam_j$ between selected pairs of teams
    (left: Quinnipiac and Colgate; right: Quinnipiac and Clarkson),
    based on the Bradley-Terry-Davidson model applied to the 2020-2021
    ECAC results, with all overtime games counted as ties.  Curves are
    as defined in \fref{f:ECBTQnCg}}
  \label{f:ECWLTQnCg}
\end{figure}

As with the standard Bradley-Terry model, we can show the marginal
posterior distributions on the differences
$\{\lpair_{ij}=\lteam_i-\lteam_j\}$ between pairs of log-strength
parameters, and we do this in \fref{f:ECWLTQnCg} for the same pairs of
teams as before.  Once again, samples drawn from the multivariate
Gaussian approximation capture the shape of that distribution well,
and samples drawn from the exact posterior using Hamiltonian Monte
Carlo are slightly different but similar.

\begin{figure}[t!]
  \centering
  \includegraphics[width=0.45\textwidth]{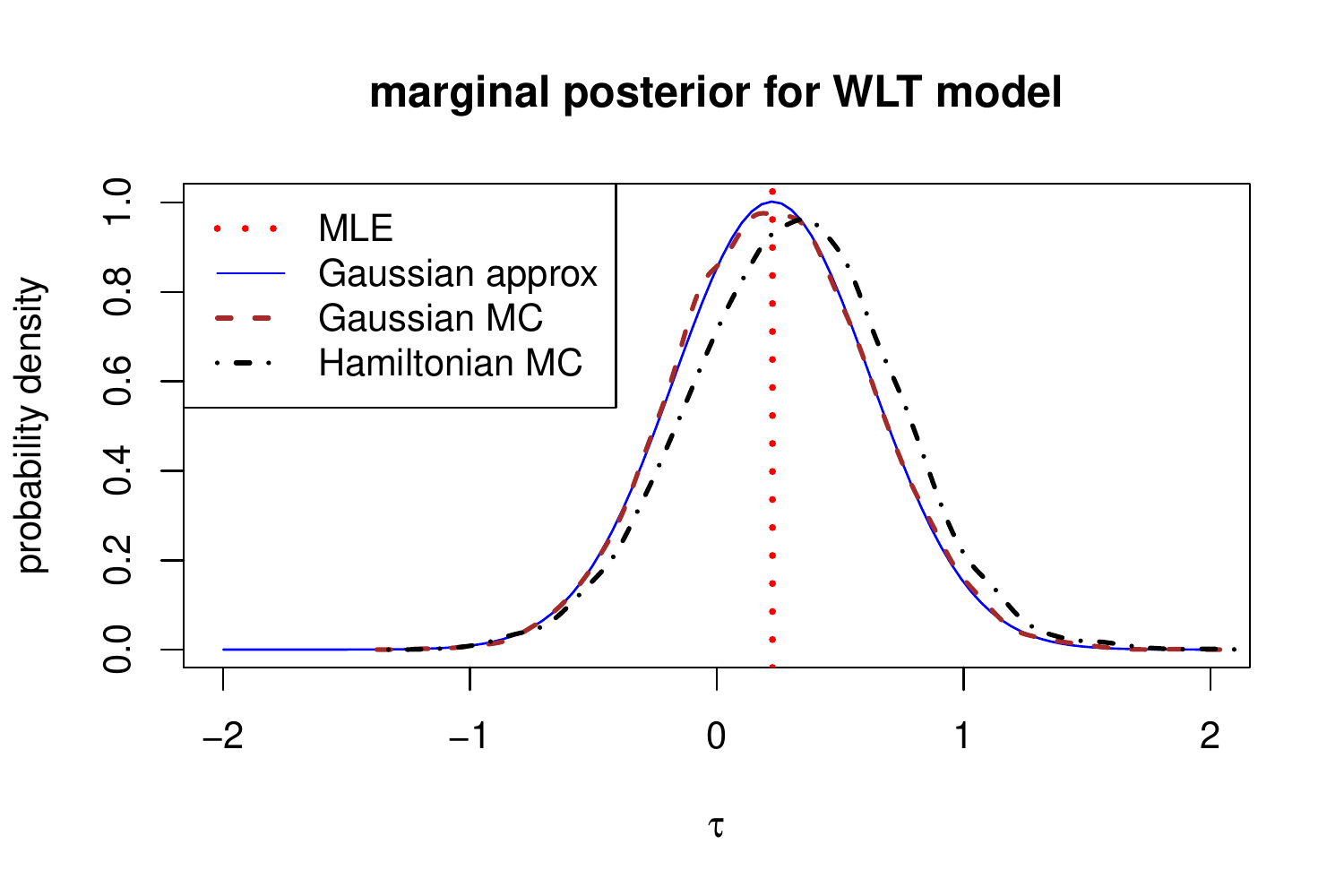}
  \includegraphics[width=0.45\textwidth]{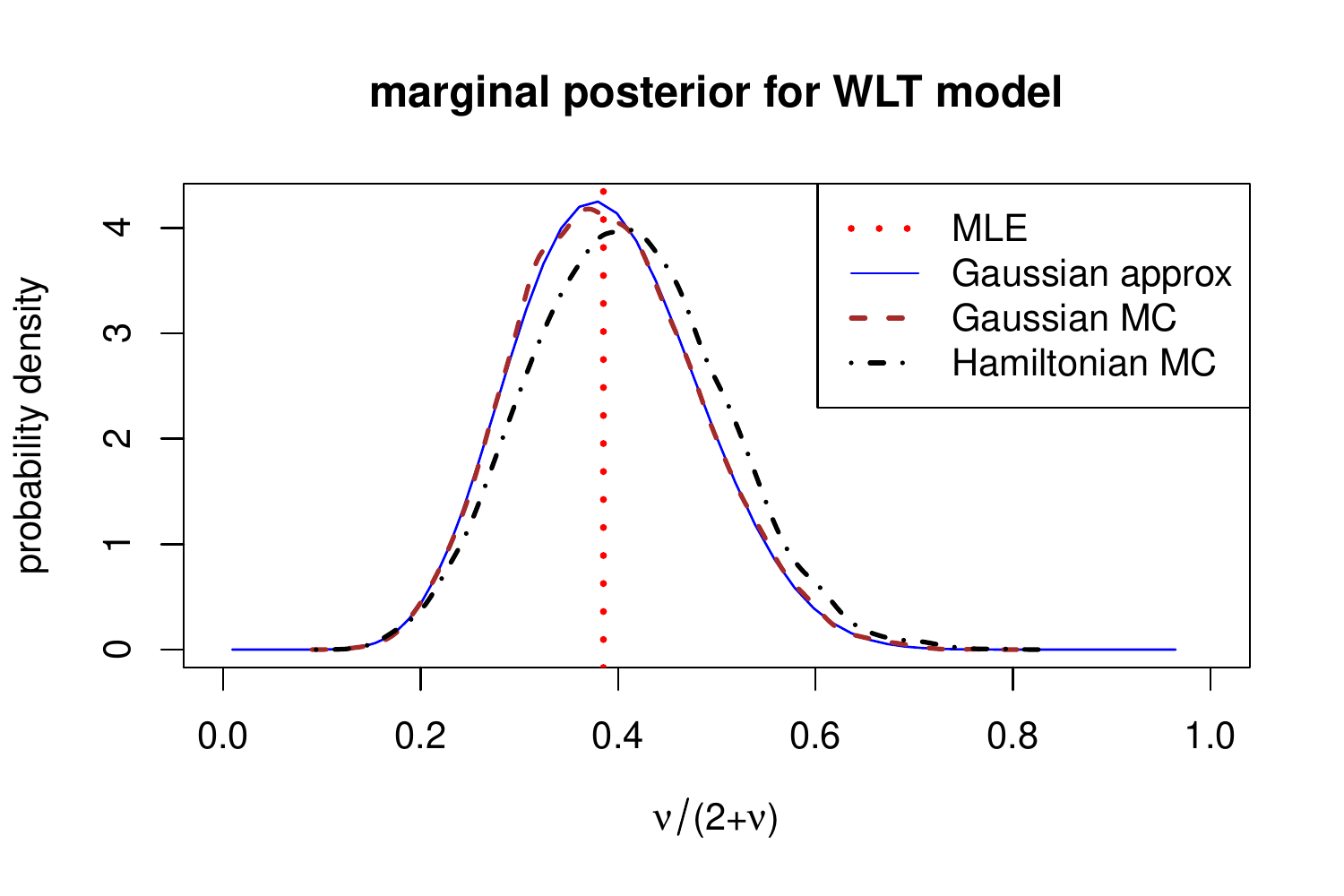}
  \caption{Posterior probability density for the log-tie parameter
    $\tau$ (left) and the associated probability
    $\frac{e^{\tau}}{2+e^{\tau}}$ (right) of a tie game between evenly
    matched teams, in the Bradley-Terry-Davidson model applied to the
    2020-2021 ECAC results, with all overtime games counted as ties.
    As in \fref{f:ECBTQnCg} and \fref{f:ECWLTQnCg}, the dashed
    vertical line is the maximum-likelihood estimate, the solid blue
    line is a Gaussian approximation to the posterior but expanding
    about the MAP point $\ML{\tau}$, and the dashed brown and dot-dash
    black lines are densty estimates, respectively constructed from a
    Monte Carlo sample from the approximate Gaussian distribution and
    from the exact distribution using Hamiltonian Monte Carlo.  Note
    that while the MLE is the maximum of the marginal posterior on
    $\tau$, the transformation of the posterior probability density
    means $\frac{e^{\ML{\tau}}}{2+e^{\ML{\tau}}}$ is not the maximum
    of the posterior on $\frac{e^{\tau}}{2+e^{\tau}}$}
  \label{f:ECWLTtau}
\end{figure}

\begin{figure}[t!]
  \centering
  \includegraphics[width=0.45\textwidth]{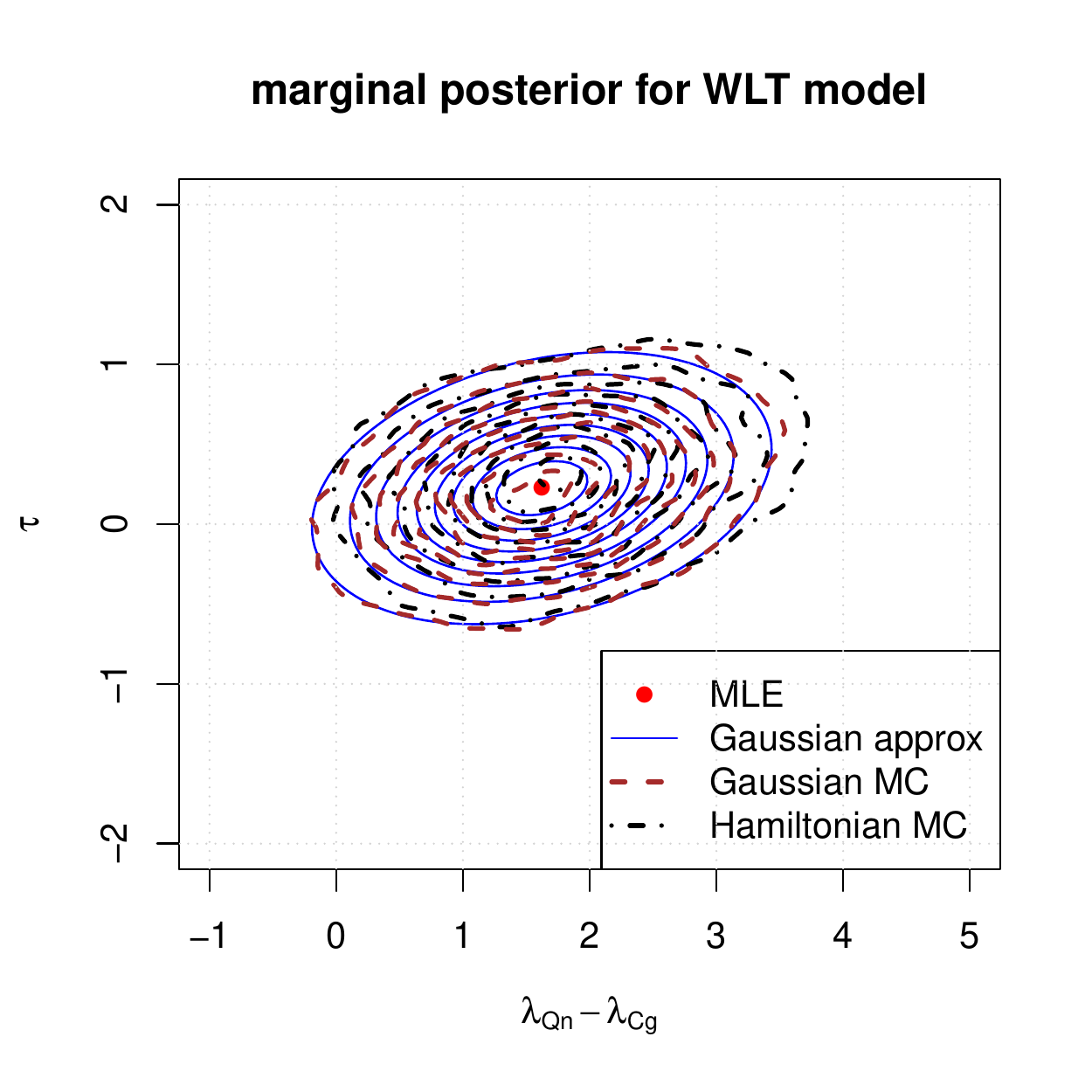}
  \includegraphics[width=0.45\textwidth]{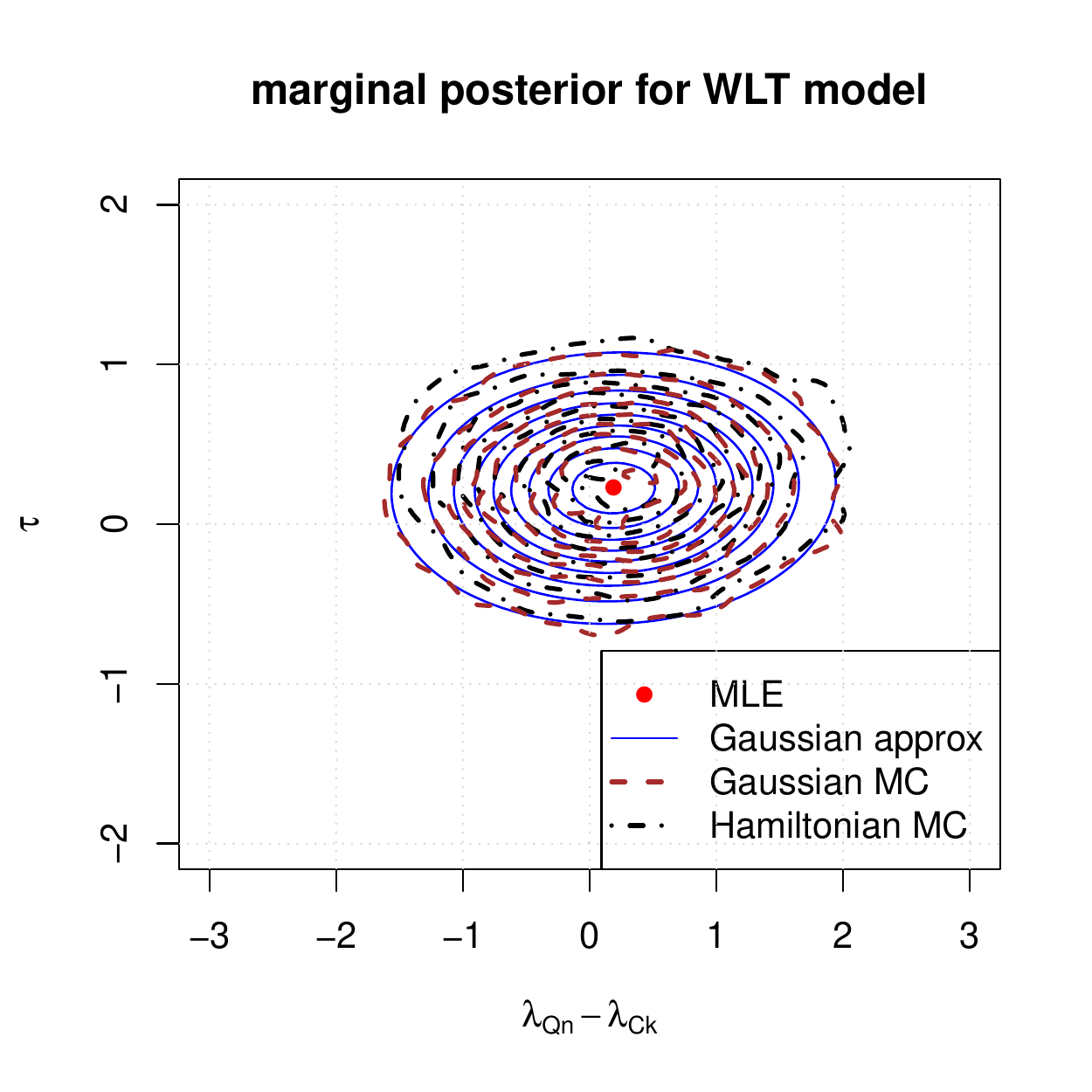}
  \caption{Contours of the joint posterior probability density of the
    log-strength differences $\lpair_{ij}=\lteam_i-\lteam_j$ shown in
    \fref{f:ECWLTQnCg} and the log-tie parameter $\tau$ shown in the
    left panel of \fref{f:ECWLTtau}.  The red circle is the MLE
    $\ML{\lpair}_{ij},\ML{\tau}$.  The solid blue curve are contours
    of the Gaussian approximation, and the dashed brown curves are
    density contours of a Monte Carlo sample drawn from that
    approximate distribution.  The dot-dashed black curves are density
    contours of a sample from the exact distribution drawn using
    Hamiltonian Monte Carlo.  As with the Bradley-Terry model, the
    exact and approximate posteriors are comparable, but differences
    are detectable beyond the level of the Monte Carlo uncertainties
    illustrated by the difference between the ``Gaussian approx'' and
    ``Gaussian MC'' contours}
  \label{f:ECWLTscat}
\end{figure}

We cannot convert $\lpair_{ij}$ directly into a probability, however,
since probabilities depend on the log-tie parameter $\tau$ as well.
In \fref{f:ECWLTtau} we plot the marginal posterior on $\tau$.  The
parameter $\tau$ can be transformed into a probability
$\frac{\nu}{2+\nu}$ where ($\nu=e^{\tau}$) for a game between
evenly-matched teams to be tied, and we plot the posterior for this as
well.  Finally, in \fref{f:ECWLTscat} we illustrate the joint marginal
posterior in $\lpair_{ij}$ and $\tau$ for our selected pairs of teams.

\begin{figure}[t!]
  \centering
  \includegraphics[width=0.45\textwidth]{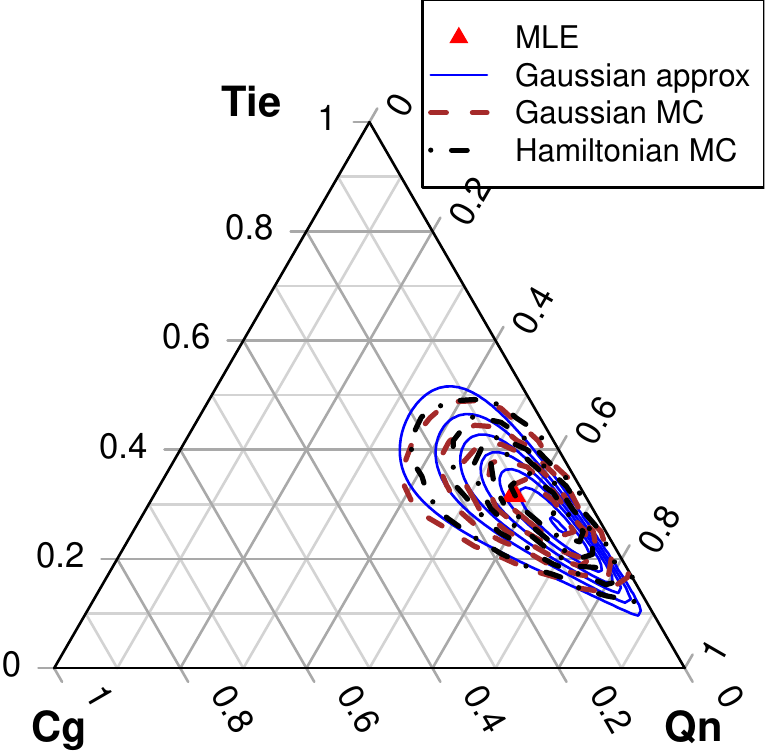}
  \includegraphics[width=0.45\textwidth]{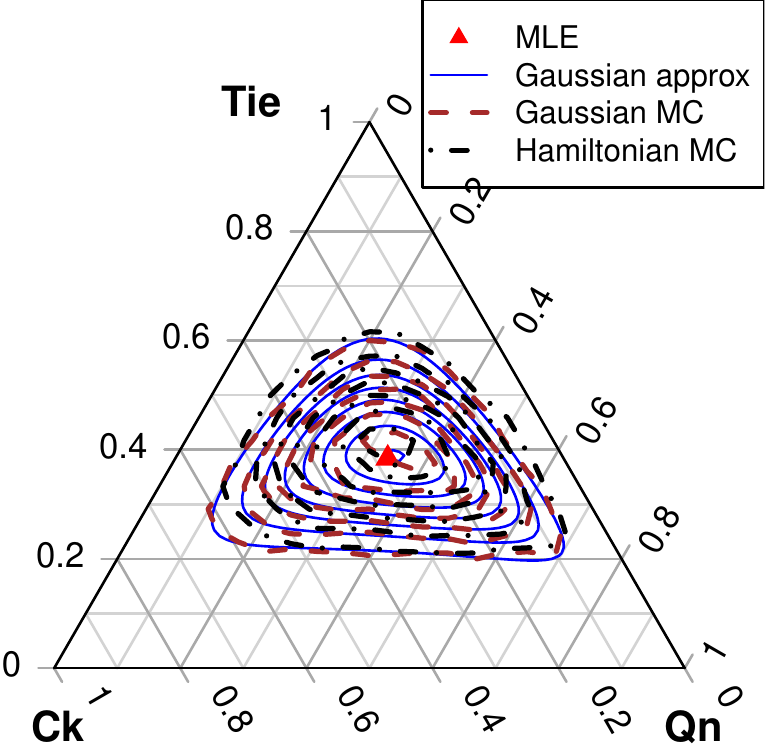}
  \caption{Ternary plots illustrating the joint posterior on
    $\ppair^{\text{W}}_{ij}$, $\ppair^{\text{T}}_{ij}$, and
    $\ppair^{\text{L}}_{ij}$, based on the Bradley-Terry-Davidson
    model applied to the 2020-2021 ECAC results, with all overtime
    games counted as ties.  The horizontal gridlines correspond to
    lines of constant $\ppair^{\text{T}}_{ij}$, with
    $\ppair^{\text{T}}_{ij}=1$ labelled as ``Tie''; the diagonal
    gridlines correspond to lines of constant $\ppair^{\text{W}}_{ij}$
    or $\ppair^{\text{L}}_{ij}$, with $\ppair^{\text{W}}_{ij}=1$
    labelled with the abbreviation for team $i$ (``Qn'' for Quinnipiac
    in both cases) and $\ppair^{\text{L}}_{ij}=1$ labelled with the
    abbreviation for team $j$ (``Cg'' for Colgate and ``Ck'' for
    Clarkson).  The red triangle is the maximum likelihood point
    $\ML{\ppair}^{I}_{ij}$.  Note that for a given set of game
    results, the maximum likelihood point for all pairs of teams will
    lie along a one-dimensional curve in the Ternary plot. since, for
    a fixed $\ML{\tau}$, the maximum-likelihood probabilities are
    functions of the single value $\ML{\lpair}_{ij}$.  The three sets
    of contours are as defined in \fref{f:ECWLTscat}.  Note that the
    MLE is no longer the maximum of the posterior probability density
    after tranforming parameters from $\lpair_{ij},\tau$ to
    $\ppair^{\text{W}}_{ij}$, $\ppair^{\text{T}}_{ij}$, and
    $\ppair^{\text{L}}_{ij}=1-\ppair^{\text{W}}_{ij}-\ppair^{\text{T}}_{ij}$}
  \label{f:ECTern}
\end{figure}

To illustrate the posterior on the probabilities
$\{\ppair^{I}_{ij}|I=\text{W},\text{T},\text{L}\}$ for a pair of
teams, we note that the constraint $\sum_I\ppair^{I}_{ij}=1$ means
that the space is actually two dimensional.  The natural visualization
for the behavior of three quantities which sum to one is a ternary
plot, and we contour plot density estimates of the posterior and its Gaussian
approximation in \fref{f:ECTern}, along with the maximum likelihood
estimates $\{\ML{\ppair}^{I}_{ij}\}$.

\subsection{ECAC: Bradley-Terry-like Model with Overtime/Shootout Results}

\label{s:ECAC}

\begin{figure}[t!]
  \centering
  \includegraphics[width=0.45\textwidth]{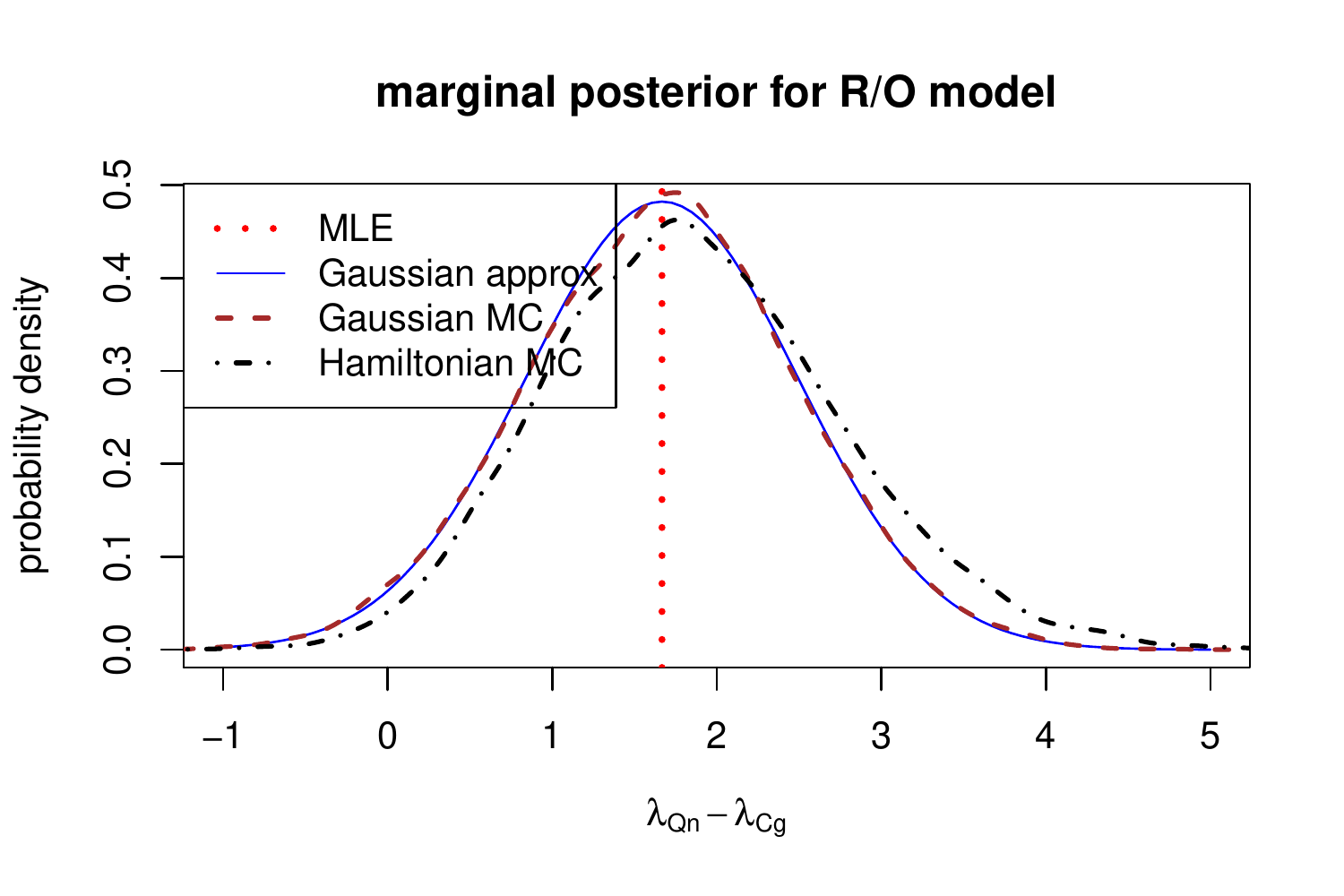}
  \includegraphics[width=0.45\textwidth]{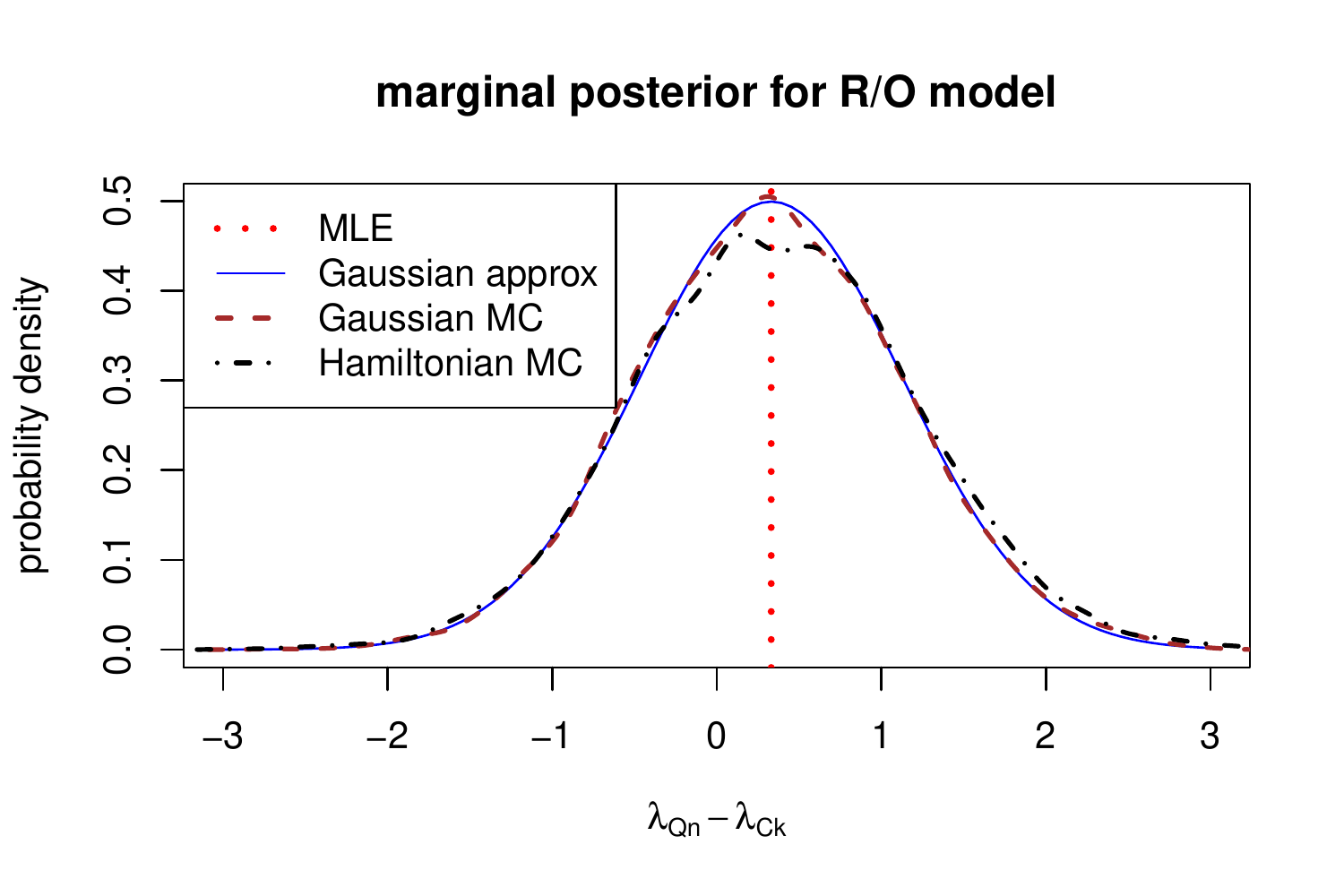}
  \caption{Posterior probability density for difference in
    log-strengths $\lteam_i-\lteam_j$ between selected pairs of teams
    (left: Quinnipiac and Colgate; right: Quinnipiac and Clarkson),
    based on the Bradley-Terry-like model with four game outcomes
    applied to the 2020-2021 ECAC results.  Curves are
    as defined in \fref{f:ECBTQnCg}.}
  \label{f:ECQnCg}
\end{figure}

\begin{figure}[t!]
  \centering
  \includegraphics[width=0.45\textwidth]{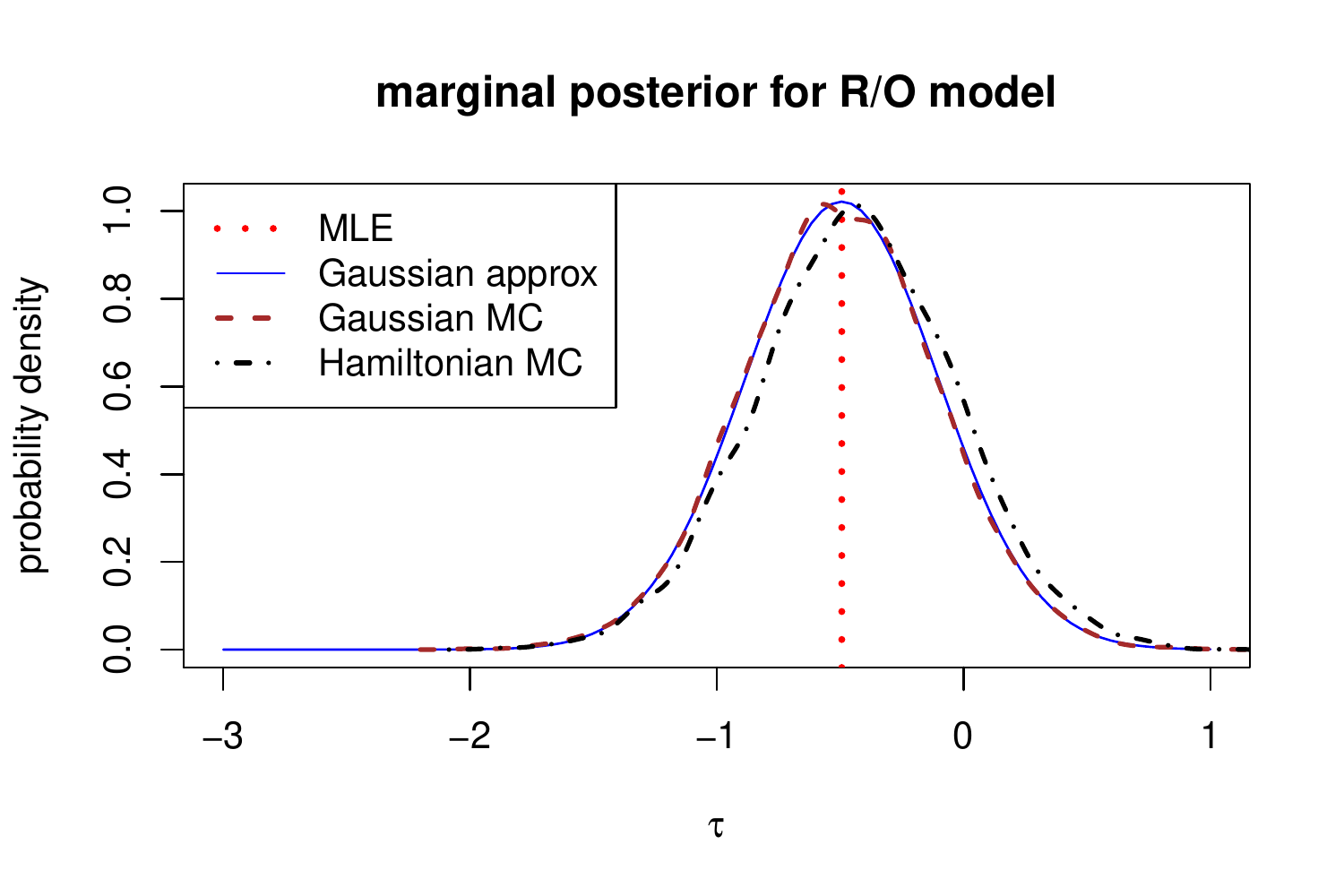}
  \includegraphics[width=0.45\textwidth]{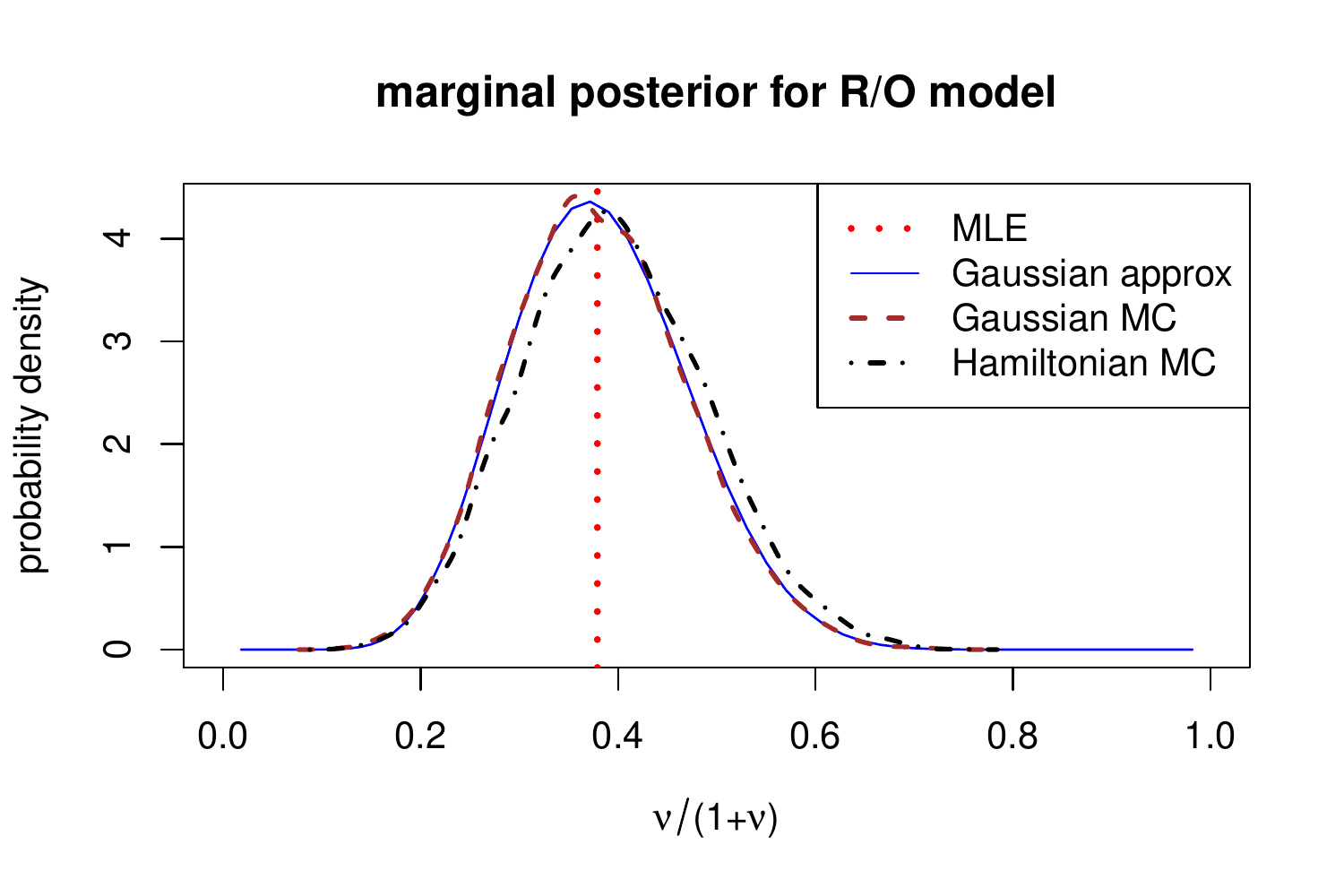}
  \caption{Posterior probability density for the log-overtime
    parameter $\tau$ (left) and the associated probability
    $\frac{e^{\tau}}{1+e^{\tau}}$ (right) of an overtime game between
    evenly matched teams, in the Bradley-Terry-like model with four
    game outcomes applied to the 2020-2021 ECAC results.  Curves are
    as defined in \fref{f:ECWLTtau}.  Note that the posterior on the
    overtime probability is very similar to that for the tie
    probability in the right panel of \fref{f:ECWLTtau}.  This is not
    surprising since the two calculations are based on different
    interpretations of the same set of game results, and the ``ties''
    used to generate \fref{f:ECWLTtau} are just the overtime games in
    the current computation.  The estimates on $\tau$ appear different
    in the two models, but that is mostly because $\nu=e^{\tau}$ is a
    measure of the probability of each type of overtime result
    compared to each type of regulation result, and there are two
    overtime results in this model and only one in the
    Bradley-Terry-Davidson model with ties}
  \label{f:ECtau}
\end{figure}

\begin{figure}[t!]
  \centering
  \includegraphics[width=0.45\textwidth]{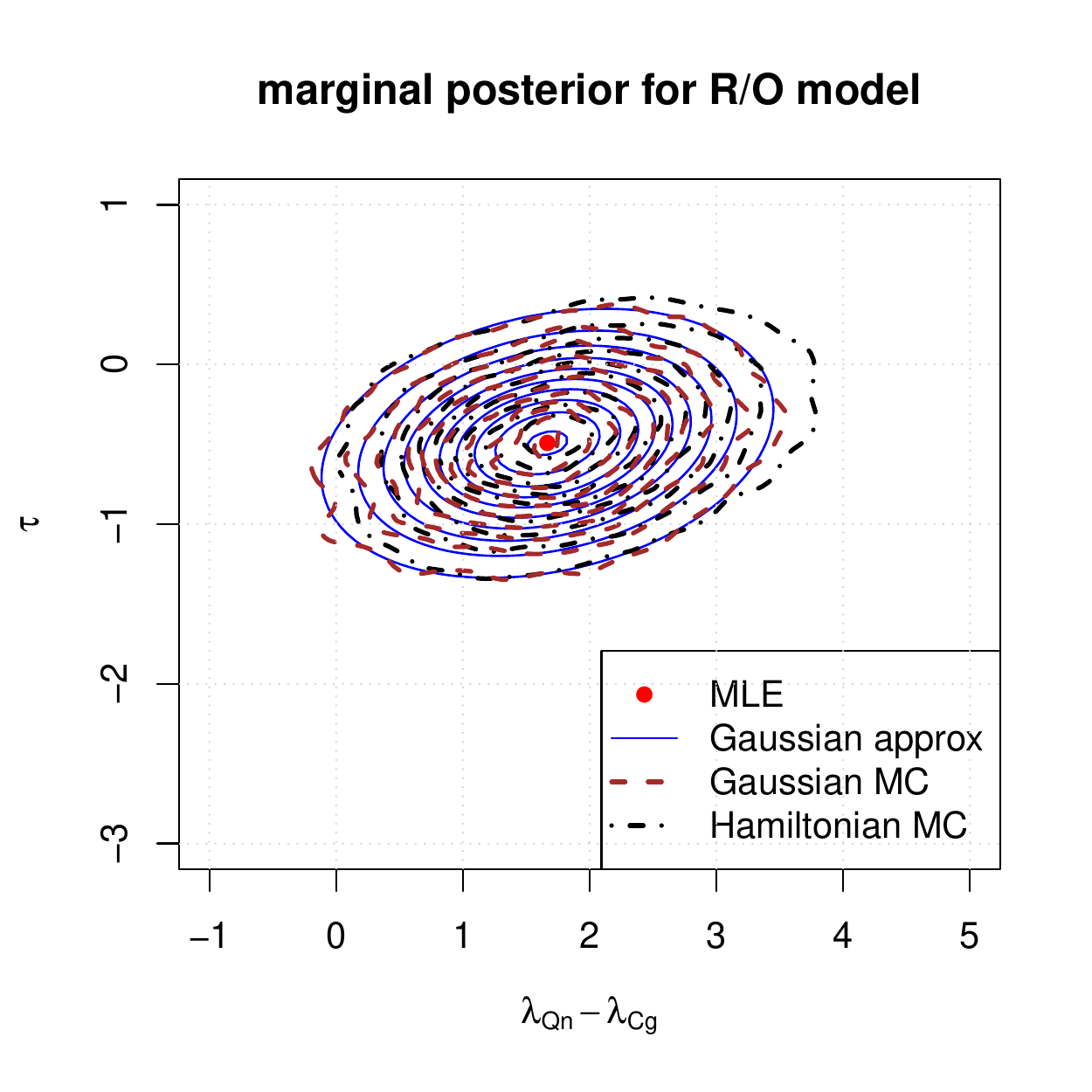}
  \includegraphics[width=0.45\textwidth]{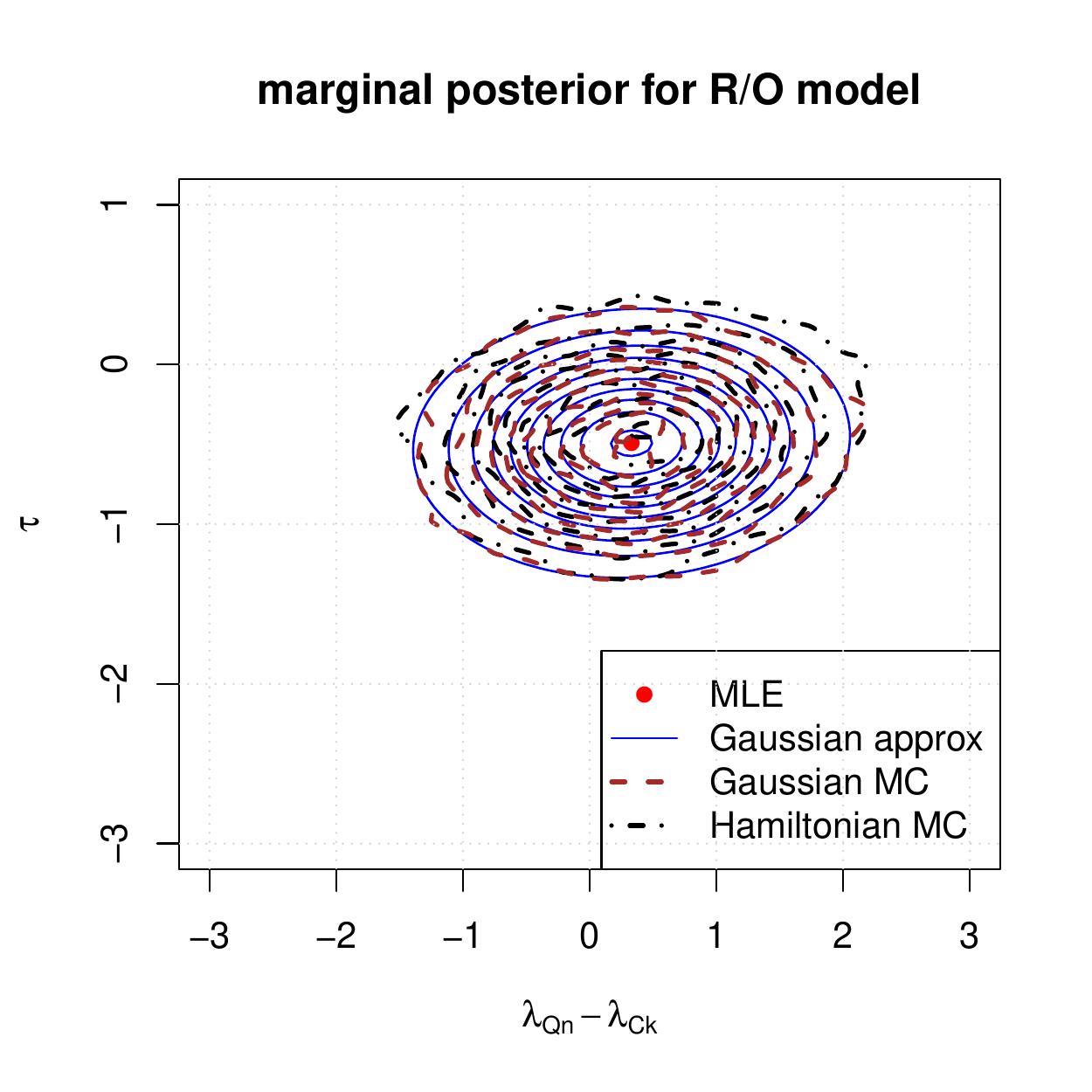}
  \caption{Samples from the joint posterior probability density of the
    log-strength differences $\lpair_{ij}=\lteam_i-\lteam_j$ shown in
    \fref{f:ECQnCg} and the log-overtime parameter shown in the left
    panel of \fref{f:ECtau}.  Contours are as defined in
    \fref{f:ECWLTscat}.}
  \label{f:ECscat}
\end{figure}

\begin{figure}[t!]
  \centering
  \includegraphics[width=0.45\textwidth]{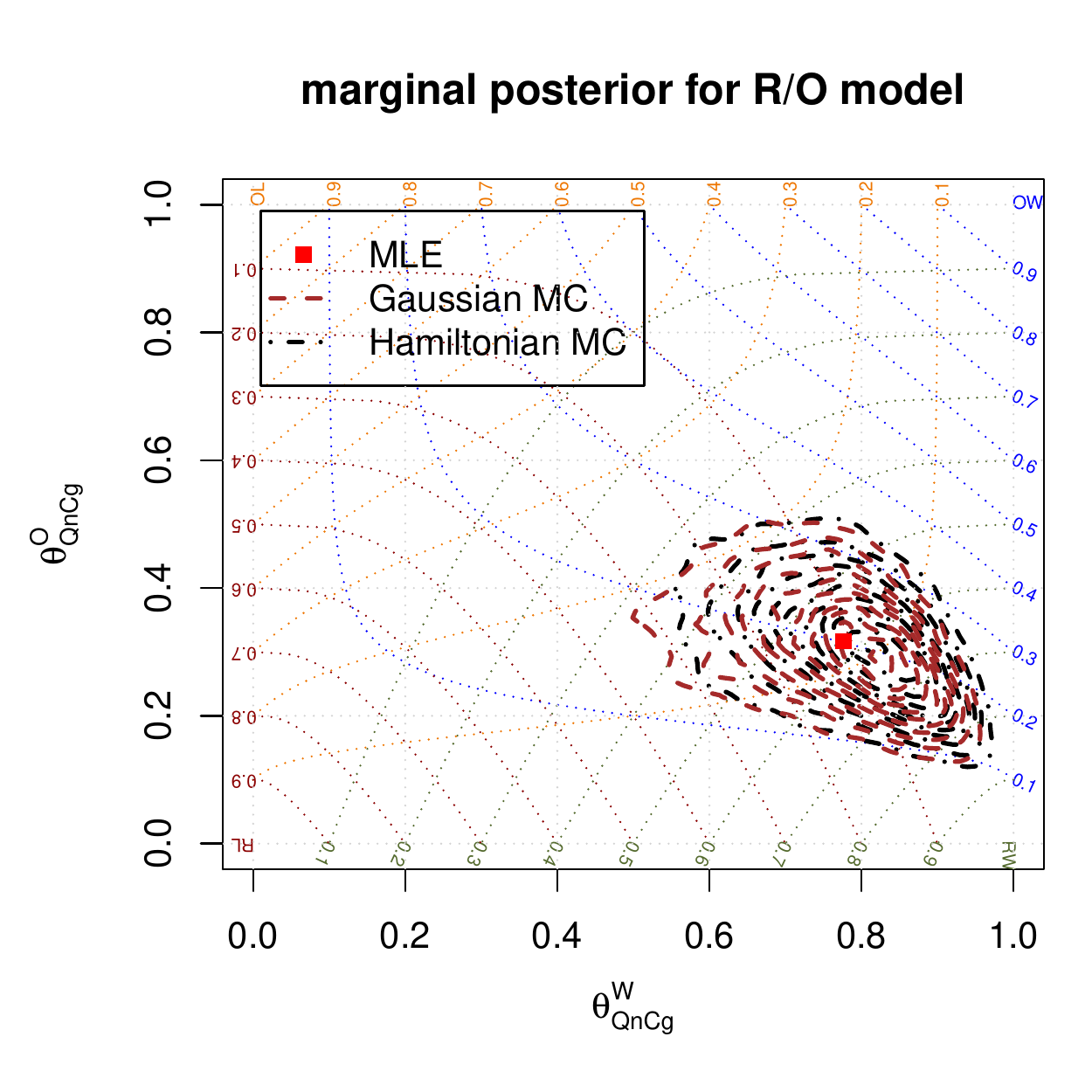}
  \includegraphics[width=0.45\textwidth]{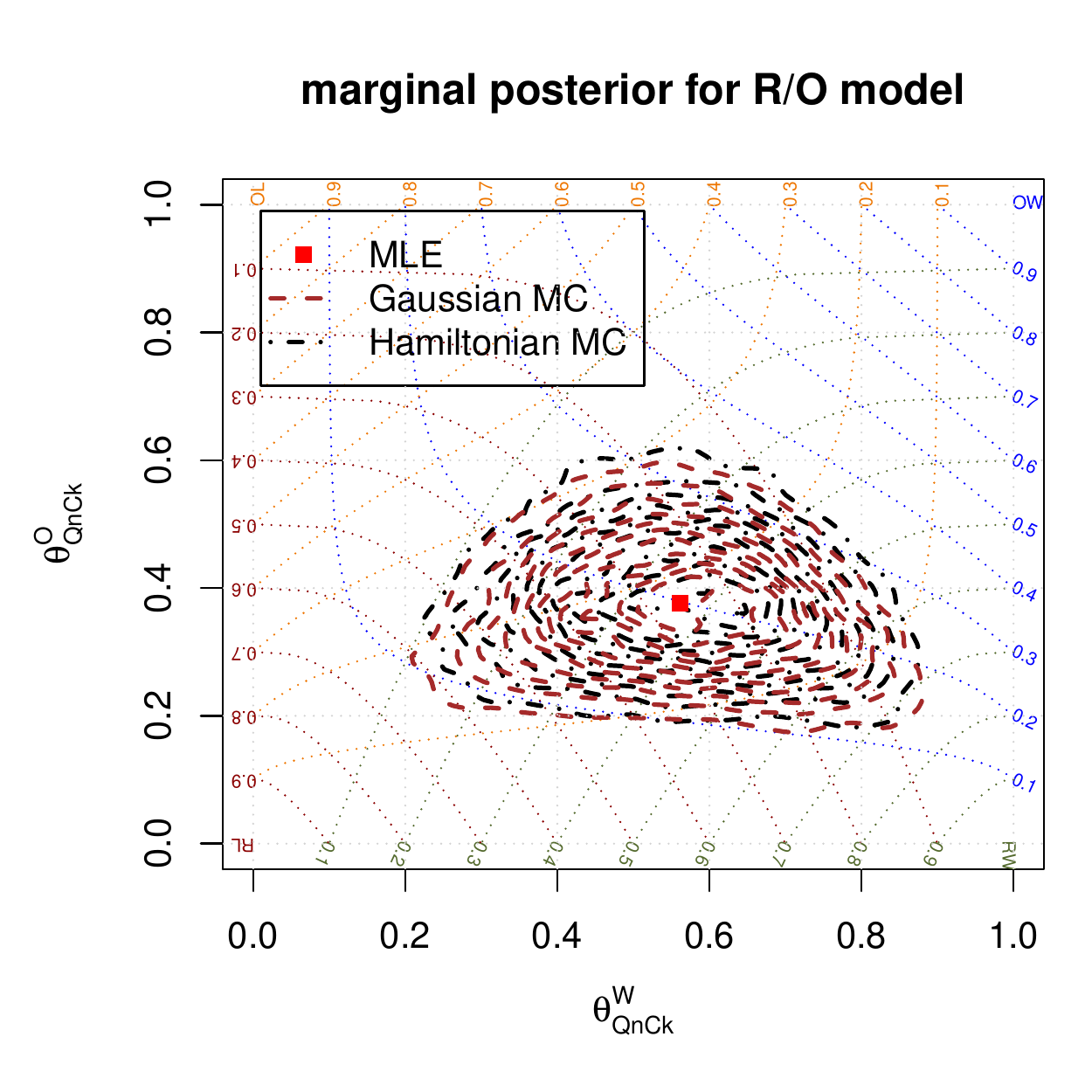}
  \caption{Density contours from the joint posterior probability
    distribution of
    $\ppair^{\text{W}}_{ij}=\ppair^{\text{RW}}_{ij}+\ppair^{\text{OW}}_{ij}$
    $\ppair^{\text{O}}_{ij}=\ppair^{\text{OW}}_{ij}+\ppair^{\text{OL}}_{ij}$,
    transformed from the joint probability on $\lpair_{ij}$ and $\tau$
    shown in \fref{f:ECscat}.  Each point on this plot can be
    converted into a set of probabilitues $\{\ppair^{I}_{ij}|I\}$
    using the relations
    $\frac{\ppair^{\text{OW}}_{ij}}{\ppair^{\text{OL}}_{ij}} =
    \left(\frac{\ppair^{\text{RW}}_{ij}}{\ppair^{\text{RL}}_{ij}}\right)^{1/3}$
    and
    $\ppair^{\text{RW}}_{ij} + \ppair^{\text{OW}}_{ij} +
    \ppair^{\text{OL}}_{ij} + \ppair^{\text{RL}}_{ij} = 1$.  The red
    square is the maximum-likelihood estimate
    $\ML{\ppair}^{\text{W}}_{ij},\ML{\ppair}^{\text{O}}_{ij}$.  The
    dashed brown curves are density contours of a Monte Carlo sample
    drawn from the Gaussian approximation to the posterior
    distribution on $\{\lteam_i\}$ and $\tau$.  The dot-dashed black
    curves are density contours of a sample from the exact
    distribution drawn using Hamiltonian Monte Carlo.  As usual, while
    the MLE is the maximum a posteriori point in the parameters
    $\lpair_{ij}$ and $\tau$, it is not so in the parameters shown
    here due to the transformation of the posterior probability
    density.}
  \label{f:ECthetascat}
\end{figure}

Having developed the mechanisms to characterize the posterior
distribution for the Bradley-Terry-Davidson model with three outcomes
(win, tie, and loss), we apply similar analogues for the model with
four outcomes: regulation win (RW), overtime/shootout win (OW),
overtime/shootout loss (OL), and regulation loss (RL), now applied to
the full 2020-2021 ECAC results shown in \tref{t:ECdata}.  As before,
there is a log-strength parameter $\lteam_i$ for each team, and $\tau$
is now the log of a parameter associated with overtime results.  We
show the maximum likelihood estimates in \tref{t:ECMLE} along with the
probabilities $\{\ML{\ppair}^{\text{RW}}_{ij}\}$ for a regulation win
$\{\ML{\ppair}^{\text{OW}}_{ij}\}$ for an overtime/shootout win in
contests between pairs of teams.  In \tref{t:ECGauss} we show the
parameters of the Gaussian approximation to the posterior.
\begin{table}[t!]
  \centering
  \caption{The maximum likelihood estimates for a Bradley-Terry-like
    model with four game outcomes applied to the 2020-2021 ECAC
    results.  The maximum likelihood estimates $\{\ML{\lteam_i}\}$ and
    $\ML{\tau}$ of the log-strengths and log overtime parameter are
    used to compute the estimated probability
    $\ML{\ppair}^{\text{RW}}_{ij}$ for a regulation win and
    $\ML{\ppair}^{\text{OW}}_{ij}$ for an overtime/shootout win
    between each pair of teams.  Note that the estimated probability
    of a game between evenly-matched teams to got to overtime  is
    $\frac{e^{\ML{\tau}}}{1+e^{\ML{\tau}}}={\ECOTprob}$, and it is
    lower the more different the two teams' strengths are.}
  \begin{tabular}{l|c|cccc}
& & \multicolumn{4}{|c}{$\ML{\ppair}^{\text{RW}}_{ij}$ ($\ML{\ppair}^{\text{OW}}_{ij}$)} \\ 
Team $i$ & $\ML{\lteam_i}$ & Cg & Ck & Qn & SL \\ 
 \hline 
Cg & $-0.74$
 & ---  & $0.14$ ($0.13$) & $0.11$ ($0.12$) & $0.32$ ($0.19$) \\ 
Ck & $0.60$
 & $0.53$ ($0.21$) & ---  & $0.26$ ($0.18$) & $0.53$ ($0.20$) \\ 
Qn & $0.93$
 & $0.57$ ($0.20$) & $0.36$ ($0.20$) & ---  & $0.58$ ($0.20$) \\ 
SL & $-0.79$
 & $0.30$ ($0.19$) & $0.13$ ($0.13$) & $0.10$ ($0.11$) & ---  \\ 
 \hline 
\multicolumn{6}{c}{$\ML{\tau}=-0.49$}\end{tabular}
   \label{t:ECMLE}
\end{table}
\begin{table}[t!]
  \centering
  \caption{The parameters of the the Gaussian approximation to the
    posterior distribution for the Bradley-Terry-like model with four
    game outcomes applied to the 2020-2021 ECAC results.  In addition
    to the log-strength parameters $\{\lteam_i\}$, there are
    uncertainties and correlations associated with the log-overtime
    parameter $\tau$.}
  \begin{tabular}{l|cc|ccccc}
& & & \multicolumn{4}{|c}{$\rho_{ij}=\Sigma_{ij}/\sqrt{\Sigma_{ii}\Sigma_{jj}}$} \\ 
Team $i$ & $\ML{\lteam}_i$ & $\sqrt{\Sigma_{ii}}$ & Cg & Ck & Qn & SL & $\tau$ \\ 
 \hline 
Cg & $-0.74$ & $0.48$
 & $1.00$ & $-0.34$ & $-0.41$ & $-0.17$ & $-0.19$ \\ 
Ck & $0.60$ & $0.54$
 & $-0.34$ & $1.00$ & $-0.17$ & $-0.52$ & $0.14$ \\ 
Qn & $0.93$ & $0.50$
 & $-0.41$ & $-0.17$ & $1.00$ & $-0.38$ & $0.23$ \\ 
SL & $-0.79$ & $0.56$
 & $-0.17$ & $-0.52$ & $-0.38$ & $1.00$ & $-0.18$ \\ 
$\tau$ & $-0.49$ & $0.39$
 & $-0.19$ & $0.14$ & $0.23$ & $-0.18$ & $1.00$ \\ 
\end{tabular}
   \label{t:ECGauss}
\end{table}
As in the Bradley-Terry-Davidson model, we can plot the marginal
parameters for the differences $\{\lpair_{ij}=\lteam_i-\lteam_j\}$
between pairs of log-strength parameters (\fref{f:ECQnCg}), the
log-overtime parameter $\tau$ or equivalently the probability
$\frac{\nu}{1+\nu}$ where ($\nu=e^{\tau}$) for a game between
evenly-matched teams to go to overtime (\fref{f:ECtau}), and the joint
marginal posterior in $\lpair_{ij}$ and $\tau$ for our selected pairs
of teams (\fref{f:ECscat}).
The posterior distribution on the probabilities
$\{\ppair^{I}_{ij}|I=\text{RW},\text{OW},\text{OL},\text{RW}\}$ is
more difficult to visualize, because we have four probabilities which
sum to 1, so the posterior can be thought of as defined on the
interior of a tetrahedron, which is an example of an Aitchison simplex
\cite{Aitchison1982}.  However, since all four probabilities are
determined by two parameters $\lpair_{ij}$ and $\tau$, they must lie
on a (curved) two-dimensional subsurface of the simplex, defined by
the constraint
$\frac{\ppair^{\text{OW}}_{ij}}{\ppair^{\text{OL}}_{ij}} =
\left(\frac{\ppair^{\text{RW}}_{ij}}{\ppair^{\text{RL}}_{ij}}\right)^{1/3}$
as well as
$\ppair^{\text{RW}}_{ij} + \ppair^{\text{OW}}_{ij} +
\ppair^{\text{OL}}_{ij} + \ppair^{\text{RL}}_{ij} = 1$.  In
\fref{f:ECthetascat} we illustrate one possibility for a
two-dimensional plot of the marginal posterior on $\ppair^{I}_{ij}$,
by plotting posterior density contours in
$\ppair^{\text{W}}_{ij}=\ppair^{\text{RW}}_{ij}+\ppair^{\text{OW}}_{ij}$
(the probability of any sort of a win) and
$\ppair^{\text{O}}_{ij}=\ppair^{\text{OW}}_{ij}+\ppair^{\text{OL}}_{ij}$
(the probability of an overtime result).  This has the conceptual
advantage that each side of the square corresponds to an edge of the
tetrahedrom, and each vertex of the square corresponds to a vertex of
the tetrahedron, at which $\ppair^{I}_{ij}=1$ for some result $I$.
However, the conversion of a point
$\ppair^{\text{W}}_{ij},\ppair^{\text{O}}_{ij}$ into $\ppair^{I}_{ij}$
is nontrivial and cannot be written in closed form, so further
investigation of methods of presenting the posterior is called for.

\section{Discussion and Conclusions}
\label{s:conclusions}

We have defined a generalization of Davidson's extension to the
Bradley-Terry outcome that handles the set of game outcomes currently
distinguished in ice hockey: regulation wins, overtime/shootout wins,
overtime/shootout losses, and regulation losses.  We've explicitly
computed maximum likelihood estimates, constructed a Gaussian
approximation to the likelihood, and drawn posterior samples directly
from the Gaussian approximation or from the exact posterior using the
Hamiltonian Monte Carlo method implemented in Stan.  For the data sets
examined, the Gaussian approximation produced similar (but slightly
different) results to the exact posterior.  The differences in
log-team strengths were qualitatively similar among the original
Bradley-Terry model (\sref{s:ECACBT}), the Bradley-Terry-Davidson
model with ties (\sref{s:ECACWLT}), and the new model including
regulation and overtime/shootout results (\sref{s:ECAC}), when applied
to the same set of results (albeit with overtime/shootout results
interpreted differently).  However, these computations are not meant
to determine a ``best'' model, but to illustrate the capabilities of
the algorithm.  (By definition, we consider the appropriate model to be the
one that corresponds to how the league actually assigns values to the
results of games in the standings.)

We now wish to discuss some limitations of the work
to date, and possible approaches to address them: the use of an
improper non-informative Haldane prior, the choice of the parameters
$\{p_{I}\}$ in the probability model, and the application of the model
to predict the outcomes of playoff games, which may not be played
under the same conditions with overtime and shootouts.

First, for simplicity, we worked with a non-informative Haldane prior
which was uniform in the log-parameters $\{\lteam_i\}$ and $\tau$, so
that the posterior probability distribution in those variables was
proportional to the likelihood.  There are a number of options for
normalizable priors on the distribution of log-strengths
$\{\lteam_i\}$ in the Bradley-Terry model (see \cite{Whelan2017} for a
discussion), of which two promising options are a Gaussian prior or a
generalized logistic prior \cite{Phelan2017,Whelan2019}, each of which
has a hyperprior which can be fixed to previous seasons' data or
estimated in a hierarchical model as in \cite{Phelan2017}.  Similar
options suggest themselves for the prior on the log-overtime parameter
$\tau$, although the situation is somewhat different in that $\tau$
has a meaningful origin, so one has to consider a possible location
parameter.  In particular, it's not clear whether the most natural
``origin'' for $\nu=e^{\tau}$ is $1$, $2$, or something else.

Second, we made something of an arbitrary choice by setting
$p_{\text{OW}}=\frac{2}{3}$ and $p_{\text{OL}}=\frac{1}{3}$.  In the
Bradley-Terry-Davidson model with ties, the requirement that
$p_{\text{T}}=p_{-\text{T}}=1-p_{\text{T}}$ means
$p_{\text{T}}=\frac{1}{2}$ is the only option, as there is only one
zero-point system in the three-outcome model.  With four outcomes,
however, $p_{\text{OW}}=\frac{2}{3}$ is a choice.  This choice was of
course informed by the point system used for the standings, so that
the maximum likelihood equations would enforce that the expected
number of points for each team equals its actual number.  Other point
systems are possible, however.  In an earlier experiment with
shootouts the Central Collegiate Hockey Association awarded 5 points
for a win in regulation or overtime, 3 for a shootout win, 2 for a
shootout loss, and 0 points for a loss in regulation or overtime, so
analysis of that season might have used $p_{\text{SW}}=\frac{3}{5}$
and $p_{\text{SL}}=\frac{2}{5}$.  Similarly, the NCAA, for tournament
selection purposes, considers a win in 3-on-3 overtime worth $0.55$ of
a win, and treats games decided in a shootout as a tie.  Capturing
this in a model would require two parameters in addition to the
team-strengths: one for overtime games and one for ties, and would
have parameters like $p_{\text{RW}}=1$, $p_{\text{OTW}}=0.55$,
$p_{\text{SO}}=0.50$, $p_{\text{OTL}}=0.45$, and $p_{\text{RL}}=0$.
One avenue for future investigation would be to define an extended
model in which the unconstrained values of $\{p_{I}\}$ are treated as
additional parameters to be estimated from the data.  For instance, in the
four-outcome model, $p_{\text{OW}}$ could be treated as a
parameter with prior support on the interval
$\frac{1}{2}<p_{\text{OW}}<1$.

Finally, the model has assumed all games are played under the same
conditions, with 3-on-3 overtimes and shootouts.  However, in a number
of hockey leagues, playoffs and other postseason games are played to
conclusion with overtimes played under the same set of rules with a
full squad on the ice, and shootouts are not possible, To produce
probabilities for such a game, one would have to decide what
probability to assign to a win or a loss.  The natural model is
probably to use $\theta^{POW}_{ij}=\frac{\pi_i}{\pi_i+\pi_j}$, i.e.,
the conditional probability of winning a game given that it's not
decided in (3-on-3) overtime or a shootout.  Likewise if any playoff
games are included in the results used for inference, their
contribution to the likelihood would need to be adjusted.

\appendix

\section{Stan Model}
\label{s:stanmodel}

Here we show the Stan model implementing the family of Bradley-Terry-like
models described in this paper.  The generalization allows a
single Stan dynamic shared object (DSO)\cite{Stan} to be used for all
three models.  This is computationally convenient, because compiling the
the DSO is often the most time-consuming part of a Stan simulation.

\begin{verbatim}
data {
  int<lower=1> nteams;
  int<lower=1> nres;
  int<lower=0> n_ttR[nteams,nteams,nres];
  int<lower=0,upper=1> o_R[nres];
  real p_R[nres];
}
parameters {
  real omega_t[nteams-1];
  real tau;
}
model {
  int n_tt[nteams,nteams];
  real gamma_tt[nteams,nteams];
  real denom_tt[nteams,nteams];
  int sumo;
  vector[nres] gamma_ttR[nteams,nteams];
  vector[nres] theta_ttR[nteams,nteams];
  sumo = 0;
  for (R in 1:nres) {
    sumo += o_R[R];
  }
  // Hack to keep tau from going crazy if the model does not have ties
  if (sumo == 0) {
    target += std_normal_lpdf(tau);
  }
  for (i in 1:(nteams-1)) {
    for (j in (i+1):nteams) {
      n_tt[i,j] = 0;
      for (R in 1:nres) {
        n_tt[i,j] += n_ttR[i,j,R];
      }
      if ( n_tt[i,j]>0 ) {
        gamma_tt[i,j] = 0;
        for (k in i:(j-1)) {
          gamma_tt[i,j] += omega_t[k];
        }
        for (R in 1:nres) {
          gamma_ttR[i,j,R] = o_R[R] * tau + p_R[R] * gamma_tt[i,j];
        }
// Unfortunately only supported in Stan 2.24 and above
//         n_ttR[i,j] ~ multinomial_logit(gamma_ttR[i,j]);
        denom_tt[i,j] = 0;
        for (R in 1:nres) {
          denom_tt[i,j] += exp(gamma_ttR[i,j,R]);
        }
        for (R in 1:nres) {
          theta_ttR[i,j,R] = exp(gamma_ttR[i,j,R]) / denom_tt[i,j];
        }
        n_ttR[i,j] ~ multinomial(theta_ttR[i,j]);
      }
    }
  }
}
\end{verbatim}

\bibliographystyle{acm}

\begin{acknowledgment}
  JTW wishes to thank Adam Wodon, Kenneth Butler, Gabriel Phelan, the
  attendees of the UP-STAT conferences, and the members of the
  Schwerpunkt Stochastik at Goethe University, Frankfurt am Main, for
  useful discussions.  Game results for the
  computations in this paper were collected from \\
  \texttt{https://www.collegehockeynews.com/} and
  \texttt{https://www.flashscore.com/}
\end{acknowledgment}
 
\end{document}